\documentclass[11pt]{cernrep}
\usepackage{graphicx}
\usepackage{here}
\begin{document}
 \title{BEYOND THE STANDARD MODEL}
\author{Esteban Roulet}
\institute{CONICET -- Centro At\'omico Bariloche, Av. Bustillo km. 9.5, 8400,
  Bariloche, Argentina}
\maketitle
\begin{abstract}
The successes and shortcomings of the Standard Model are reviewed, with emphasis
on the reasons motivating the need to extend it. The basic elements of
 grand unification and supersymmetry are described, exploring their
phenomenological  implications for gauge coupling unification, proton
decay, fermion masses, 
neutrino physics, collider signatures, dark matter, rare decays and 
anomalous magnetic moments. The forthcoming  generation of experiments will
certainly expose these ideas to several essential tests.
\end{abstract}

\section{THE STANDARD MODEL}

The experimental success of the $SU(3)\times SU(2)\times U(1)$ Standard Model (SM)
of the strong and electroweak forces can be considered as 
the triumph of the gauge symmetry
principle to describe  particle interactions.
As we now briefly summarize, there are many different facts which have to be
taken into account when searching for a deeper underlying theory, and these
are:

\subsection{THE GOOD:}
The major success of  the SM is that it 
accounts for essentially all present
accelerator results. In particular:
\begin{itemize}
\item The most accurately known quantity in particle physics 
is the magnetic
  moment of the electron, with its `anomalous' part being  $a_e\equiv
(g_e-2)/2$, i.e. the normalized 
difference in the gyromagnetic ratio $g_e$ with respect to the classical Dirac
value $g_e=2$ (where the magnetic moment $\mu$ is related to the
particle spin $s$ through $\mu\equiv gs(e/2m)$). The experimental and
theoretical values are respectively \cite{RPP}
\begin{equation}
a_e=\left\{\matrix{(115965218.7\pm 0.4)\times 10^{-11}\ \ {\rm Exp.} \cr
(115965214.0\pm 2.8)\times 10^{-11}\ \ {\rm Th.}}\right. .
\end{equation}

The theoretical expression results from the 
 computation of the loop corrections within the SM
up to order ${\cal O}(\alpha^4)$, including diagrams such as those depicted in
 Figure~1. The one loop photon correction is the well known Schwinger term.
 Its prediction was actually one of the first major successes
 of QED, proving that radiative corrections could be made meaningful through
 renormalization and were indeed measurable. Higher order contributions include terms
 such as that involving the photon self energy in diagram (1.c) and the
 photon--photon scattering appearing in diagram (1.d). These two are actually
 the major sources of uncertainties for the magnetic moments of heavier leptons
 ($\mu$ or $\tau$), but for the electron the theoretical error is actually
 dominated by the uncertainty in the direct 
measurement of $\alpha$ using the Quantum Hall Effect.  Indeed, one may use
 the theoretical expression for $a_e$ (as a fourth order polynomial in
 $\alpha$) to infer a `theoretically based' value for the electromagnetic
 coupling, which has a smaller error 
than the one obtained  from direct measurements,
 and is $\alpha^{-1}=137.03599993(52)$. 
This of course is the value at low energies, and running it to the scale of
 the $Z$ boson, where for instance LEP measurements are done, one gets in the
modified minimal subtraction ($\overline{MS}$) scheme
\begin{equation}
\alpha^{-1}(M_Z)=\left({\alpha\over 1-\Delta\alpha}\right)^{-1}=127.934\pm
0.027,
\end{equation}
where $\Delta\alpha$ encodes the effects of the radiative corrections.

\begin{figure}
\begin{center}
\includegraphics[width=10.5cm]{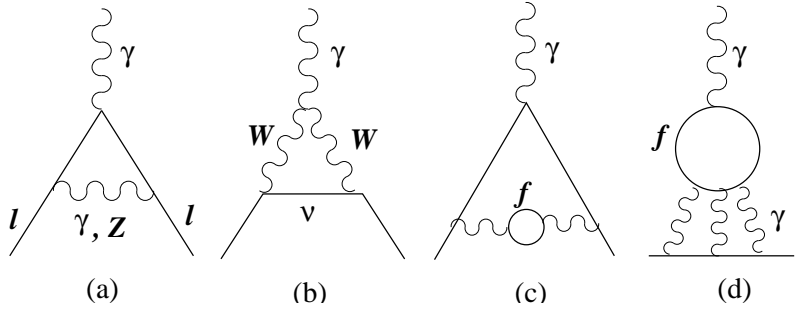} 
\caption{Some loop contributions to the anomalous magnetic moments of
charged leptons.}
\end{center}
\end{figure}

\item The weak interactions are based on the spontaneous breaking $SU(2)_L\times
  U(1)_Y\to U(1)_{em}$ (with associated coupling constants $g$, $g'$ and $e$
  respectively) induced by the vacuum expectation value of the neutral
  component of the Higgs doublet field $\langle H^0\rangle=246$~GeV. 
Assuming that the left handed fermions belong to SU(2) doublets while the right
handed chiralities to SU(2) singlets ensures that the
charge  current
couplings of the $W$ boson violate parity maximally, and this gives rise to the
well established $V-A$
theory. At low energies, $W$-boson mediated processes have an effective
coupling with strength given by the Fermi constant,
$G_F/\sqrt{2}=g^2/(8M_W^2)$, and the
measured muon lifetime leads to the value $G_F=1.16637(1)\times
10^{-5}$~GeV$^{-2}$.
Regarding the neutral currents, the $\gamma$ and $Z$ bosons are obtained
rotating by an angle $\theta_W$ the neutral SU(2) gauge boson $W^3$ and
the $U(1)_Y$ boson $B$, i.e.\footnote{We denote $s_W\equiv
\sin\theta_W$ and $c_W\equiv \cos\theta_W$.}
\begin{equation}
\pmatrix{A_\mu\cr Z_\mu}=\pmatrix{c_W & s_W\cr -s_W & c_W}\pmatrix{B_\mu\cr
  W^3_\mu}.
\end{equation}
The  photon is the gauge boson associated to the unbroken $U(1)_{em}$ (and
hence remains massless) provided that
\begin{equation}
{\rm tg}\theta_W={g'\over g},
\label{tgw}
\end{equation}
and its coupling to fermions is vectorial,  
having strength $e\equiv gs_W$ and being
proportional to the fermion charge $Q_f=T_3(f)+Y(f)$.
Regarding the fermion couplings to the $Z$ boson, they can be written as
($g/c_W)\gamma_\mu(g_V^f-g_A^f\gamma_5)$, with the vectorial part
being $g_V^f=T_3(f)-2Q_fs_W^2$ and the axial vector piece being
$g_A^f=T_3(f)$. The resonant production of $Z$ bosons at LEPI allowed
to test these couplings and 
to accurately measure the $Z$ boson mass
from  the observed line-shape, resulting in  $M_Z=91.1872\pm
0.0021$~GeV. 
One can also obtain directly the masses
of the gauge bosons from the spontaneously broken electroweak
Lagrangian and one obtains the relation $M_W=c_W M_Z$. Clearly this
relation and Eq.~(\ref{tgw}) cannot both hold at the loop level, since
the couplings in Eq.~(\ref{tgw}) run and are hence scale dependent. 
This leads to different definitions for the weak mixing angle, with
the on-shell value being
\begin{equation}
\sin^2\theta_W\equiv 1-{M_W^2\over M_Z^2}=0.22302\pm 0.00040,
\end{equation}
while the (scale dependent) $\overline{MS}$ one being
\begin{equation}
\sin^2\hat\theta_W(\mu)\equiv {g'^2(\mu)\over g^2(\mu)+g'^2(\mu)},
\end{equation}
with fits to electroweak observables leading to
$\sin^2\hat\theta_W(M_Z)=0.23117\pm 0.00016$. The $\overline{MS}$ definition of
the mixing
angle is the most appropriate one for the study of the running of
gauge couplings, and in particular to confront unification predictions.
Combining the large amount of electroweak observables, including LEPI Z
resonance  cross sections, widths and asymmetries, Tevatron and LEPII  W-boson
mass measurements, and also neutrino scattering processes, one can test the
effects of radiative corrections, which
 are sensitive to the virtual effects of the top quark and Higgs
boson.
From these one obtains favored
ranges for the
top and Higgs masses, which are $m_t=174.1^{+9.7}_{-7.6}$~GeV and
$m_H=86^{+48}_{-32}$~GeV \cite{la01}, 
in remarkable agreement with the top mass measured
at the Tevatron, $m_t=174.3\pm 5.1$~GeV, and suggesting the presence
of a light Higgs boson\footnote{Some hints in favor of a Higgs 
mass of 115~GeV have even been suggested by LEPII data.}. Hence, one can say that 
also the electroweak sector has 
been tested at the loop level.

\item The strong interactions are described by Quantum Chromo Dynamics, i.e. 
by the unbroken non-abelian gauge
  theory SU(3). Due to the gluon's self interactions it has the property of
  being asymptotically free, while at low energies the coupling constant
  becomes large and the theory should then account for the confinement of
  quarks into colorless hadrons. The running of the strong
  coupling has been tested extracting $\alpha_s$ from experiments performed at
  different energies, such as the measured $\tau$ lepton widths, deep inelastic
  scattering, Upsilon decays and $e^+e^-$ colliders at different center of
  mass energies and up to $\sim 200$~GeV. In particular, one has
  $\alpha_s(M_Z)=0.119\pm 0.003$.

\item Besides the gauge sector, a crucial ingredient of the SM is the family
  structure. The first generation of fermions consists of
$$
L=\pmatrix{\nu_e\cr e}_L\ \ ,\ \  e_R\ \ ,\ \  Q=\pmatrix{u\cr d}_L\ \ ,\ \ 
 u_R\ \ ,\ \  d_R $$
and this pattern is replicated two more times to lead to the three fermion
families. This number of three is nicely consistent with the number of
massless neutrinos coupling to the Z-boson inferred from the invisible Z
width, $N_\nu=2.994\pm 0.012$, and also with the possibility of having CP
violation in the quark sector through a non-trivial phase in the Cabibbo
Kobayashi Maskawa matrix $V_{CKM}$. The CKM phase can account for the CP violating
effects observed in the Kaon system ($\varepsilon$ and $\varepsilon'/\varepsilon$) or
B system (time dependent asymmetries in $B^0\to J/\Psi K_S$).

The two quark states contained in the  doublet $Q$ are directly coupled among
them through a W boson, and hence constitute the so-called flavor eigenstates. The
mass eigenstate quarks are instead in general a mixture of flavor
eigenstates belonging to different families. Adopting for convenience the up
type ($u,\ c$ and $t$) flavor eigenstates to coincide with the corresponding
mass eigenstates, the down type mass eigenstates ($d'_i=(d',s',b'$)) are
just related to the flavor eigenstates through the CKM matrix, i.e.
\begin{equation}
\pmatrix{d'\cr s'\cr b'}=V_{CKM}\pmatrix{d\cr s\cr b}.
\end{equation}
The unitarity of this matrix, which is ultimately due to the family structure
of the model, ensures that the couplings of the Z-boson to the down type mass
eigenstates is also flavor diagonal, since they turn out to be proportional to
$(V^\dagger V)_{ij}=\delta_{ij}$. This is the basis of the Glashow
Illiopoulos and Maiani (GIM) mechanism \cite{gl70}, 
which forbids tree level flavor
changing neutral currents (FCNC) and supresses the loop mediated ones, which
are non-zero only due to the mass differences between different quarks. This
naturally accounts for the smallness of e.g. the $\Delta S=1$ decay $K^0_L\to
\mu^+\mu^-$ (with $BR\simeq 7\times 10^{-9}$) or the $\Delta S=2$ neutral kaon mass
difference ($\Delta m_K/m_K\simeq 7\times 10^{-15}$), as illustrated in
Figure~\ref{gim}. The value of $\Delta m_K$ actually provided the first
indication of the correct mass for the charm quark \cite{ga74}, so that as in the
previously mentioned example of the top mass, it is important to keep in mind
that in these two cases virtual processes were sensitive to the effects of
particles prior to their direct production at accelerators. One should then not
be too skeptical when looking for virtual effects as a way to discover
 new particles predicted in extensions of the SM (e.g. supersymmetric or GUT ones).

\begin{figure}
\begin{center}
\includegraphics[width=10.cm]{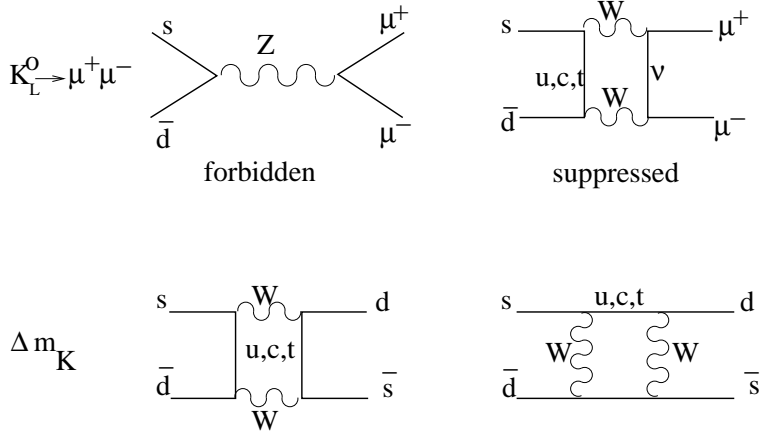} 
\caption{GIM suppressed processes $K^0_L\to \mu^+\mu^-$ and
contributions to the $\Delta S=2$ neutral Kaon mass difference.}
\label{gim}
\end{center}
\end{figure}

Due to the large hierarchy $m_t\gg m_c$, in the flavor changing processes involving
the $b$ quark, such as $b\to s\gamma$, the GIM suppression is not very
effective, leading to sizeable decay rates.  One has indeed 
\begin{equation}
BR(B\to X_s\gamma)=\left\{ \matrix{(3.72\pm 0.33)\times 10^{-4}\ Th.\cr 
(2.96\pm 0.35)\times 10^{-4}\ Exp.}\right.,
\label{bsgam}
\end{equation}
and the reduced suppression in this process makes it also an interesting probe
for new physics.

\end{itemize}

\subsection{THE BAD:}
Together with the many good things of the SM, there are several bad aspects
where it fails to provide an adequate solution. These are:
\begin{itemize}
\item There is by now quite solid evidence supporting the existence of
  non-zero neutrino masses, coming from the explanation of the atmospheric and
  solar neutrino problems in terms of neutrino oscillations. The SM with just
  left handed neutrinos and a Higgs doublet is unable to provide a non-zero
  neutrino mass. On the other hand, the masslessness of the neutrinos in
  the SM is not related to any deep symmetry principle (unlike the
  masslessness of the photon which is linked to gauge invariance), and hence
  it is quite common that extensions of the SM give rise to massive neutrinos.
The simplest way to get massive neutrinos would be to
  introduce the right-handed neutrino states $\nu_R$, and obtain a Dirac
  mass term by means of a Yukawa coupling $-{\cal
    L}_Y=\lambda_\nu\bar\nu_L\nu_RH+h.c.$. However, for the resulting masses to
  be sufficiently small (below the eV range) the Yukawa coupling would have to be quite
  small ($\lambda_\nu<10^{-11}$), and this doesn't seem very natural. All
  other attempts to give masses to the neutrinos require to extend the SM in
  more radical ways. 

\item The SM suffers from the so-called strong CP problem, which is the fact
  that a Lorentz invariant term in the SM Lagrangian of the form
\begin{equation}
{\cal L}_{QCD}\supset {\theta_{QCD}\over 32\pi^2}G_{\mu\nu}\tilde
G^{\mu\nu},
\end{equation}
is consistent with the gauge symmetries of the model and there is
hence no reason to omit it (where $\tilde{G}^{\mu\nu}\equiv
\epsilon^{\mu\nu\rho\sigma}G_{\rho\sigma}/2$ is the dual of the gluon
field strength $G_{\mu\nu}$). This term is however CP violating and
induces a contribution to the neutron electric dipole moment $d_n\simeq
 5\times 10^{-16} \theta_{QCD}$e~cm. The experimental
upper bounds on $d_n$ require then that $\theta_{QCD}<10^{-10}$, and the smallness
of this parameter has no natural explanation within the SM. One possibility is
to extend the Higgs sector and introduce an axial global U(1) symmetry, the
so-called Peccei Quinn symmetry \cite{pe77}, which gets broken spontaneously leading to
the appearance of a pseudo-Goldstone boson, the axion. The anomaly in the
$U(1)_{PQ}$ leads to the axion coupling to $G\tilde G$, and the same dynamics
of the theory has the effect of producing a cancellation in 
the total contribution to $d_n$.

\item An important theoretical difficulty of the SM is the hierarchy problem,
  related to the fact that loop corrections to the scalar (Higgs) masses are
  quadratically divergent, and hence in a regularized theory with an
  ultraviolet cutoff $\Lambda$ one gets contributions at one loop $\delta
  m_H^2\propto(1 /16\pi^2)\Lambda^2$, with the different contributions
(see Figure~\ref{natural}) being proportional to $\lambda$ (the Higgs quartic
self-coupling), $g^2$ (the squared gauge coupling) or $-\lambda_f^2$
(the squared Yukawa coupling, with a minus sign due to the fermionic
nature of the loop). If one thinks of the cutoff as the scale
  where a more fundamental (and less divergent) theory enters to play a role,
  it would be hard to understand how things conspire to cancel the large loop
  correction (if $\Lambda\sim M_{Pl}$ or $M_{string}$), leaving a Higgs mass
  at the TeV scale. Since this would require an accurate fine-tuning, the
  situation is usually refered to as the naturalness or fine-tuning problem. The
  simplest solution for this is to appeal to supersymmetry entering
into the game  at a relatively low
  scale ($\Lambda_{SUSY}\sim {\rm TeV}$) and
enforcing the cancellation between the dangerous quadratic divergences
arising from bosonic and fermionic loops.

\begin{figure}
\begin{center}
\includegraphics[width=12.5cm]{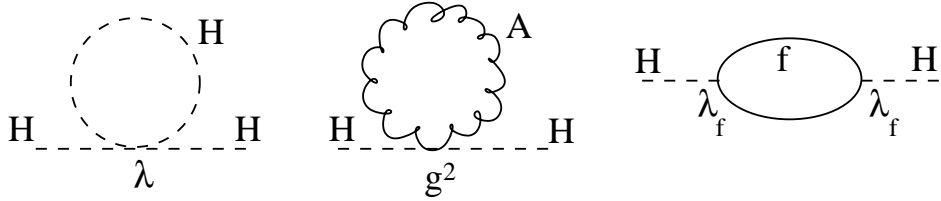} 
\caption{Quadratically divergent contribution to the scalar self-energies.}
\label{natural}
\end{center}
\end{figure}

\item The SM provides no explanation for the origin of the baryon asymmetry,
  i.e. the excess of matter over antimatter in   the Universe. Even if there
  are baryogenesis scenarios exploiting the non perturbative SM baryon number
  violation, in order for them to be successful new sources of CP violation
  and extended Higgs sectors are generally required.

\item The SM has no good candidate  for the dark matter which is inferred to
  contribute sizeably to the overall matter in the Universe. The favored cosmological
  model nowadays suggests indeed $\Omega_{CDM}\simeq 0.3$ and
  $\Omega_\Lambda\simeq 0.7$, and neither a Cold Dark Matter particle nor an
  explanation for the origin of a 
  cosmological constant $\Lambda$ of the required size can be found  within the
  SM. Cosmological theories of primordial inflation also require to search for
  their causes beyond the SM.

\item An even more serious drawback is that the SM makes no attempt to include
  a consistent quantum theory of gravity.

\end{itemize}

\subsection{THE UGLY:}

The search for a good theory to describe particle interactions has also an
esthetical component, and in physics beauty is generally related to
simplicity and to the fact that important things should not happen by
chance, but should  instead result from  solid underlying reasons.  
Some of the ugly things in the SM are:

\begin{itemize}

\item The model has many unrelated parameters. These include the three
  coupling constants $g_3$, $g$ and $g'$ and the Yukawa couplings, or
  equivalently three charged lepton masses, 6 quark masses, three CKM mixing
  angles and one CP violating phase. In the Higgs sector there is the Higgs
  quartic coupling and the Higgs VEV (or equivalently $M_H$ and $M_W$). There
  is also the QCD parameter $\theta_{QCD}$. This makes a total of 19 independent
  parameters. This number is further increased if we take into account the
  neutrino masses and leptonic mixings.
Clearly a theory relating the gauge couplings (unification) or explaining the
pattern of fermion mixings would be most welcome.

\item The gauge group $SU(3)\times SU(2)\times U(1)$ was just put in by hand
  to explain observations, but there is no deep principle behind that choice.

\item Assigning left-handed chiralities to SU(2) doublets and right-handed
  ones to singlets is again arbitrary. Left right symmetric models are
  believed to be more esthetic.

\item The number of generations $N_g=3$ is also unexplained.

\item The quantization of electric charge, i.e. the fact that $Q_d=Q_e/3$, is
  unexplained.

\item The cancellation of gauge anomalies in the SM happens just by
  chance. Let us now further comment into the anomaly issue.

\end{itemize}

\subsection{ANOMALIES}

Anomalies occur when a classical symmetry of the Lagrangian is violated by
quantum effects. A traditional example being the dilatation symmetry of a
massless (i.e. scale free) theory, 
which gets broken at the quantum level due to the need to
introduce the renormalization scale.
This leads to a non-vanishing trace of the energy momentum tensor, which
constitutes the so-called trace anomaly.

\begin{figure}[b]
\begin{center}
\includegraphics[width=10.5cm]{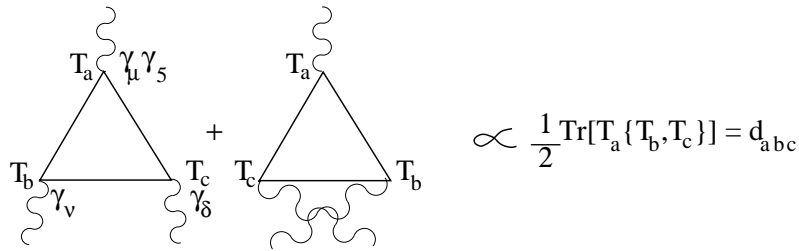} 
\caption{Diagrams contributing to the non-abelian anomaly.}
\label{dabc}
\end{center}
\end{figure}

Another example are the chiral symmetries, i.e. a symmetry under
transformations distinguishing
between left and right fermion chiralities, 
\begin{equation}
\Psi\to e^{i\theta\gamma_5}\Psi\ \ \left\{\matrix{\Psi_L=e^{-i\theta}\Psi_L\cr
    \Psi_R=e^{i\theta}\Psi_R} \right. .
\end{equation}
For a massless theory this is a good symmetry, and hence the associated
Noether current $J_\mu^5=\bar\Psi\gamma_\mu\gamma_5\Psi$ is
conserved. However, computing the one loop contribution arising from the
triangle diagrams involving one axial vector and two vector couplings (and
contracting this with the momentum incoming in the axial vector vertex, so as
to get the Fourier version of the divergence of the current), one gets
\begin{equation}
\partial^\mu J_\mu^5={g^2\over 8\pi^2}F_{\mu\nu}\tilde{F}^{\mu\nu},
\end{equation}
with $F_{\mu\nu}$ the field strength of the field coupling to the vector-like
vertex. For the
non-abelian case the resulting anomaly is proportional (as can be
seen from Figure~\ref{dabc}) to the symmetric
structure constants $d_{abc}\equiv
{\rm Tr}[T_a\{T_b,T_c\}]/2$, 
where $T_i$ are the generators in the representation of
the fermions running in the loop.

When the chiral symmetry in question is a global one, anomalies pose no
problems, and may even be a blessing. This is the case for instance in
massless QCD, where taking the limit $m_u\simeq m_d\simeq m_s\simeq 0$ the
theory acquires an $U(3)_L\times U(3)_R$ global  symmetry,
corresponding to separate unitary rotations of the left and right
quark fields $(u,d,s)_{L,R}^T$. This symmetry can be decomposed as
$SU(3)_V\times SU(3)_A\times U(1)_V\times U(1)_A$. Within QCD, 
the vectorial parts remain good symmetries, and are reflected in the
baryon multiplet degeneracies and in the baryon number conservation
respectively, while $SU(3)_A\equiv SU(3)_{L-R}$ is spontaneously
broken by the QCD chiral condensate, what is reflected in the appearance of
 Goldstone bosons which are just the meson octet. This chiral symmetry
is anomalous and the anomaly is responsible for instance for the
dominant $\pi^0$ decay ($\pi^0\to \gamma\gamma$). The $U(1)_A$ symmetry
is also anomalous, and this anomaly (combined with the non-trivial
structure of the QCD vacuum) is responsible for giving a sufficiently
large mass to the $\eta'$ meson.
When considering the electroweak interactions, since both $SU(2)_L$
and $U(1)_Y$ are chiral gauge groups, i.e. have different couplings to
left and right fermions, also the global vectorial symmetries of
baryon and lepton number become anomalous, with
\begin{equation}
\partial^\mu J_\mu^B=\partial^\mu J_\mu^L={N_f\over
32\pi^2}\left[g^2W_{\mu\nu}\tilde{W}^{\mu\nu}-g'^2Y_{\mu\nu}\tilde{Y}^{\mu\nu}\right]
,
\end{equation}
where $N_f=6$ is the number of flavors while $W_{\mu\nu}$ and
$Y_{\mu\nu}$ are the corresponding field strengths.
This anomaly implies that when topology changes in the gauge fields
occur, baryon and lepton numbers will change by an amount $\Delta
B=\Delta L =\Delta N_{CS}$, where the Chern Simons number $N_{CS}$
characterizes the amount of `winding' of the gauge fields. These
transitions are however very suppressed at low energies/temperatures,
since they are mediated by instantons, but become however very
efficient at high temperatures, where they take place through
`sphaleron' excitations, and this has profound implications for
baryogenesis theories.

Although global anomalies are welcome, when an anomaly affects a
gauge symmetry this can be a disaster, since the gauge symmetry
principle is at the basis of the formulation of the theory and is
essential for its renormalizability. Hence, if
the gauge symmetry ceases to be valid at the loop level one will
certainly be in trouble. Due to the chiral anomaly, this will in
general be the case for a chiral gauge theory, such as the one present
in the electroweak model, so that the renormalizability of the SM 
is threatened. However, it turns out that when one adds the
contributions to the anomaly coming from the leptons and from the
$N_c=3$ colors of quarks
running in the triangle diagram, one has luckily a fortuitous
cancellation, since
\begin{equation}
\sum\left(\matrix{{\rm LEPTON}\cr {\rm ANOMALY}}+N_c\times
\matrix{{\rm QUARK}\cr
{\rm ANOMALY}}\right)=0.
\end{equation}
For instance, the $SU(2)_L^2\times U(1)_Y$ contribution is
proportional to (using $Y=Q-T_3$, $\{T_a,T_b\}=\delta_{ab}/2$ and Tr$T_3=0$)
\begin{equation}
{\rm Tr}\left[\{T_a,T_b\}Y\right]={\delta_{ab}\over 2}\sum_i
Q_i\propto [Q_e+3(Q_d+Q_u)]=0.
\end{equation}
It is clear then that the SM with quarks or leptons 
alone would be anomalous, and to get a consistent electroweak theory
we need to have both of them simultaneously. 
A theory in which anomalies are absent
independently of the choice of matter representation (such as $SO(10)$
GUTs), would not have this item in the list of ugly things.

\subsection{THE ROAD TO UNIFICATION}
Many of the major advances in physics have resulted from the unified
description of aspects which were before believed to be
unrelated. This is also an important guide in the search of the
underlying theory behind the SM, trying to remedy its pitfalls
without loosing its successes. A schematic history of these unifications is
illustrated in Table~1, and we will discuss hereafter some aspects of
unification and supersymmetry  (but not deal however with other also
fundamental ideas such as extra dimensions or strings).

\begin{center}
Table 1: Brief history of unifications.
\vskip0.2cm
\begin{tabular}{|c r c l|}
\hline
{\rm Newton:} & PLANETARY MOTIONS & $\leftrightarrow$ & FALLING
APPLES\cr  
{\rm Maxwell:} &  ELECTRICITY & $\leftrightarrow$ & MAGNETISM\cr
{\rm Einstein:} & INERTIA & $\leftrightarrow$ & GRAVITY\cr
{\rm Dirac:} & PARTICLES & $\leftrightarrow$ & ANTIPARTICLES\cr
{\rm SM:} & WEAK & $\leftrightarrow$ & ELECTROMAGNETIC\cr
{\rm GUTs:} & ELECTROWEAK & $\leftrightarrow$ & STRONG \cr
{\rm SUSY:} & FERMIONS & $\leftrightarrow$ & BOSONS \cr
{\rm Extra Dimensions:} & GAUGE INTERACTIONS & $\leftrightarrow$ & GEOMETRY \cr
{\rm STRINGS:} & GAUGE INTERACTIONS & $\leftrightarrow$ & QUANTUM
GRAVITY \cr
\hline
\end{tabular}
\end{center}

\section{GRAND UNIFIED THEORIES}

The idea in Grand Unified Theories (GUTs) is to embed the SM group
$SU(3)\times SU(2)\times U(1)$ in a larger group, with the aim of
relating the different gauge couplings and also hoping that the mass
spectrum will become simpler, since the quarks and leptons become
different aspects of the same GUT `particle'. The most famous
examples of unified groups are
\begin{itemize}
\item The Pati--Salam model \cite{pa73} was the first GUT theory proposed
(1972). It was based on the $SU(4)\times SU(2)_L\times SU(2)_R$
symmetry, with the leptons being some kind of fourth quark flavor and
each family consisting of the representations
\begin{equation}
\pmatrix{u_1 & u_2 & u_3 & \nu_\ell\cr d_1 & d_2 & d_3 & \ell}_{L,R}.
\end{equation}
Putting quarks and leptons in the same $SU(4)$ multiplets lead to
proton decay processes, and the left-right symmetry required the
introduction of right handed neutrinos in each generation. Although
each group factor has an associated  gauge coupling, relations among them
can be imposed invoking discrete symmetries.

\item The smallest (i.e. ${\rm rank}=4$) 
simple gauge group containing the SM is $SU(5)$, 
and it was studied  by Georgi and Glashow in 1974 \cite{ge74}. 
Each generation is
contained in two $SU(5)$ irreducible representations, which
are\footnote{In GUTs it is always convenient to use the left handed
conjugate fields $(\Psi^c)_L$ rather than the right handed fields
$\Psi_R$, and they clearly describe the same degrees of freedom, since
$(\Psi_R)^c \equiv C\bar\Psi_R^T =(\Psi^c)_L$.} 
\begin{equation}
{\bf 10}:\pmatrix{0 & u^c_3 & -u^c_2 & u_1 & d_1\cr
-u^c_3 & 0 & u_1^c & u_2 & d_2\cr
u^c_2 & -u^c_1 & 0 & u_3 & d_3\cr
-u_1 & -u_2 & -u_3 & 0 & e^c\cr
-d_1 & -d_2 & -d_3 & -e^c &0\cr}\ \ \ ,\ \ \ \bar{\bf 5}:\pmatrix{
d_1^c\cr d_2^c \cr d_3^c \cr e\cr -\nu}.
\end{equation}

\item The orthogonal group $SO(10)$ is the next choice. It has ${\rm
rank}=5$ and can contain the previously mentioned GUT groups. The
fermions of each generation are contained in just one irreducible
representation, the ${\bf 16}$, which can be decomposed under $SU(5)$
as
${\bf 16}={\bf 10}+\bar{\bf 5}+{\bf 1}$, with the singlet state being
the right handed neutrino. 

Another attractive aspect of SO(10) is that orthogonal groups are automatically anomaly
free, since their symmetric structure constants $d_{abc}$ vanish. This is not
the case in SU(5), where one still needs the fortuitous cancellation between
the non-vanishing anomalies coming from the $\bar{\bf 5}$ and the ${\bf 10}$, which just happen
to be opposite.

\item Other larger groups which have been intensively studied are: the
exceptional group $E_6$ (with fermions in the ${\bf 27}$, which under
$SO(10)$ decomposes as ${\bf 16}+{\bf 10}+{1}$, so that many exotic
states appear in each generation); the string inspired groups
$E_8\times E_8$ (with $E_6$ being contained in one
of the $E_8$ factors)  and $SO(32)$. Large symmetry breaking chains
are required to go from these large groups down to the SM, and
many new heavy particles are left around when doing so (heavy
Higgses, new fermions, new gauge bosons such as $W_R$ or $Z'$ ones),
and these are the object of many dedicated searches at accelerators.

\end{itemize}

\subsection{GAUGE COUPLING UNIFICATION}

In the SM the three gauge couplings are different and unrelated,
i.e. $g_3\neq g\neq g'$. Moreover, the actual value of $g'$ is related
to the arbitrary normalization of the hypercharge generator. Indeed,
consider the covariant derivative in the electroweak model, which is
\begin{equation}
D_\mu=\partial_\mu -igT^a_fW^a_\mu-ig'Y_fB_\mu.
\end{equation}
The gauge fields are clearly normalized through their kinetic terms, while
the $SU(2)$ generators are normalized through the non-linear condition
\begin{equation}
{\rm Tr}(T^aT^b)={\delta_{ab}\over 2},
\label{trtatb}
\end{equation}
 and this fixes the normalization of the
coupling $g$. On the other hand, 
the $U(1)_Y$ coupling $g'$ is arbitrarily fixed by
adopting the hypercharge normalization from the relation $Y=Q-T^3$. In
an unified gauge group where all generators are similarly normalized,
the hypercharge generator will turn out to have a different
normalization than in the SM, and this modifies the expected relation
between the GUT coupling $g_{GUT}$ and $g'$. Take for instance the
case of $SU(5)$, where in the fundamental representation 
the 24 generators, normalized through the
relation (\ref{trtatb}), can be conveniently written in block form 
using the
$SU(3)$ Gellman matrices $\lambda^i$ and $SU(2)$ Pauli matrices
$\sigma^i$, as
$$
T^{i=1,...,8}={1\over 2}\pmatrix{\lambda^i & 0\cr 0 & 0}\ \ ,\ \
T^{9,10,11}={1\over 2}\pmatrix{0 & 0\cr 0 & \sigma^{1,2,3}}$$
\begin{equation}
T^{12}=\sqrt{3\over 5}Y\ \ ,\
  \ T^{13,...,24}:\ {\rm off-diagonal}
\end{equation}
We see that $T^{12}$ is proportional to the hypercharge generator
$Y={\rm
  diag}(-\frac{1}{3},-\frac{1}{3},-\frac{1}{3},\frac{1}{2},\frac{1}{2})$,
 but is now
 properly normalized. The covariant derivative of the unified group is obtained as
$D_\mu=\partial_\mu-ig_{GUT}V^a_\mu T^a=\partial_\mu-ig_{GUT}\left( 
...+W_\mu^3T^{11}+B_\mu T^{12}+...\right)$. From this we see that the
correct 
identification should be
\begin{equation}
g=g_{GUT}\ \ , \ \ g'=\sqrt{\frac{3}{5}}g_{GUT}.
\end{equation}
This clearly implies that
\begin{equation}
{\rm sin}^2\theta_W={g'^2\over g^2+g'^2}=\frac{3}{8}.
\end{equation}
This prediction, which corresponds to $s^2_W=0.375$, is clearly far from the
measured value, but this can be remedied once it is realized that it should
hold at the GUT scale, which for many reasons turns out to be quite
large ($M_{GUT}>10^{14}$~GeV), and hence the prediction is sizeably
affected by the running of the gauge couplings.

Notice that the photon should be contained among the GUT gauge bosons, and
indeed the charge generator is just given by
$Q=T^{11}+\sqrt{5/3}T^{12}$. Since the generators are traceless, this implies
that Tr$Q=0$. Considering for instance the fundamental $\bar{\bf 5}$
representation, this implies that $Q_e=3Q_d$, explaining the charge quantization
relation just from the fact that quarks and leptons are inside the same GUT
multiplets. 

\subsection{RUNNING OF GAUGE COUPLINGS}

\begin{figure}
\begin{center}
\includegraphics[width=10.5cm]{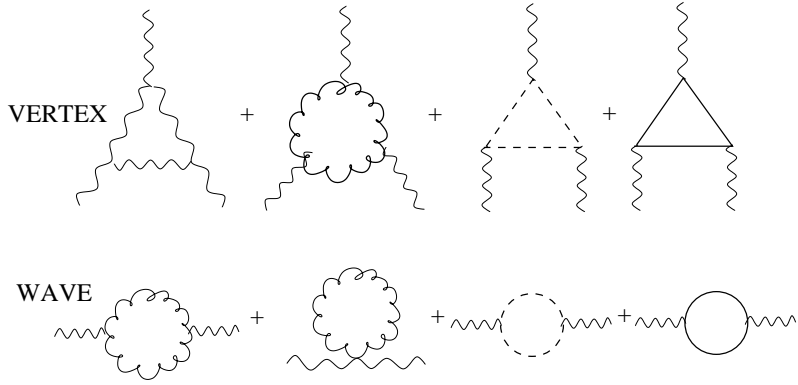} 
\caption{Diagrams determining the vertex and gauge boson wave function
renormalization, required to compute the beta functions. The scalar
lines correspond to Fadeev-Popov ghosts, and in the case of SU(2) and
$U(1)_Y$ include the Higgs bosons.}
\label{vertex}
\end{center}
\end{figure}

To obtain the running of the gauge couplings in the SM it is necessary to
consider the loop corrections to the vertex and wave functions of the gauge
bosons (depicted in Figure~\ref{vertex}), or alternatively one may
consider the gauge boson coupling to
  fermions and the fermion wave function renormalization. The result
of these two approaches is of course 
  the same, as expressed in a Slavnov Taylor identity reflecting the
  constraints imposed by the gauge symmetry.

The evolution of the
coupling constant is obtained from the solution of the 
renormalization group equations (RGE)
 d$ g/{\rm d} t=\beta$, with
$t\equiv {\rm ln}(\mu/\mu_0$) specifying the momentum scale.
For the SU(N) group one obtains the beta function 
from direct computation of the
diagrams in Figure~\ref{vertex} (e.g. they are obtained at one loop from
the residues of the $1/\varepsilon$ poles in dimensional
regularization), and they are
\begin{equation}
\beta={g^3\over 16\pi^2}\left[ -\frac{11}{3}t_2(V)+\frac{2}{3}t_2(F)+
\frac{1}{3}t_2(S)\right],
\label{beta}
\end{equation}
with  $t_2$ defined through
\begin{equation}
{\rm Tr}\left[T^aT^b\right]=t_2\delta^{ab},
\end{equation}
and where the generators are in the appropriate representation of vectors ($V$),
fermions ($F$) or scalars $(S$).
For the fields in the adjoint one has that the  Casimir is $t_2(V)=N$,
 while  for the fundamental
representation one has $t_2(F,S)=1/2$ for either Weyl fermions or complex
scalars (and twice as much for a Dirac fermion).
Hence, for QCD one has
\begin{equation}
\beta_{QCD}= {g^3\over 16\pi^2}\left[ -11+\frac{2}{3}\times \frac{1}{2}\times
  N_f\times 2\right],
\end{equation}
including the $N_f=6$ flavors of Dirac quarks coupled to the gluons.
It proves convenient to introduce the factors $b_i$ such that
\begin{equation}
{{\rm d}g_i\over {\rm d}t}\equiv {b_i\over 16\pi^2}g_i^3,
\end{equation}
and hence the solution of the RGE
 for the running coupling constant at one loop 
can be expressed as
\begin{equation}
g_i^2(t)={g_i^2(0)\over 1-(g_i^2(0)/8\pi^2)b_it}.
\label{git}
\end{equation}
For the SU(3), SU(2) and U(1) couplings of the SM one then gets (using
also that for $U(1)_Y$ one has $t_2(F,S)=Y^2$ while $t_2(V)=0$) 
\begin{equation}
b_3=-7\ \ ,\ \ b_2=-{19\over 6}\ \ ,\ \ b_1={41\over 10}.
\end{equation}

\begin{figure}
\begin{center}
\includegraphics[width=9.5cm]{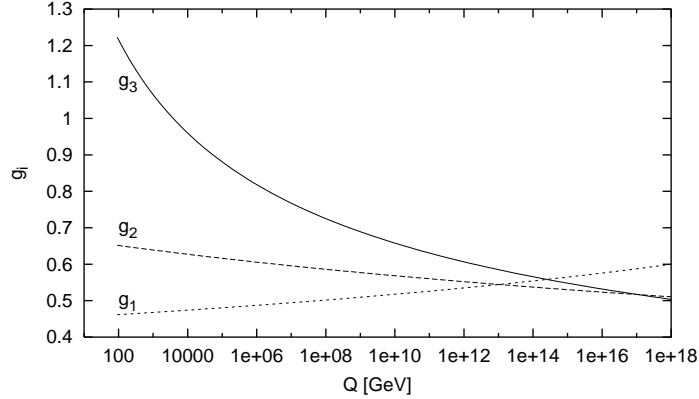} 
\caption{Running of the gauge couplings in the SM at one loop.}
\label{smrge}
\end{center}
\end{figure}

We can then start from the measured values of the coupling constants
at the scale $M_Z$, which are obtained from $\alpha_s$, $\alpha$ and
sin$^2\hat\theta_W$ through 
$g_3=\sqrt{4\pi\alpha_s}$, $g_2\equiv
g=\sqrt{4\pi\alpha}/\sin\hat\theta_W$ and
$g_1=\sqrt{5/3}g\tan\hat\theta_W$, and run them to high energies using
the RGE. The result is plotted in Figure~\ref{smrge}. We see
that the negative value of $b_3$ makes the strong coupling to become
weaker at high energies, i.e. QCD becomes asymptotically free. Also
$g_2$ becomes smaller in the SM with increasing energies, but with a
smaller slope, while $g_1$ increases with energy\footnote{In any case
the Landau pole, i.e. the energy scale 
at which $g_1$ would blow up according to Eq.~(\ref{git}), is well
beyond the Planck scale.}. We see that although the three couplings
are quite different at low energies, they have the tendency to unify
at a scale $M_X\simeq 10^{14}$--$10^{15}$~GeV. This is certainly
encouraging and may be pointing to the existence of an underlying GUT
symmetry at large scales. However, within the SM the convergence of
the gauge couplings at one scale is not perfect (and this is not
solved by including two loop contributions nor threshold effects), so
that something more will be required to get an accurate unification of the
couplings at a unique scale.

\subsection{GUTS AND FERMION MASSES}

\begin{figure}
\begin{center}
\includegraphics[width=5.cm]{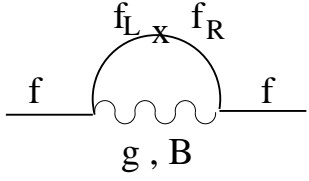} 
\caption{Diagram responsible for the running of the masses.}
\label{mrun}
\end{center}
\end{figure}

Another important issue that GUTs can address is to explain some of
the observed patterns of fermion masses. Let us consider for instance
the case of SU(5) in some detail. 
The breaking of SU(5) down to the SM is performed by the VEV of an
adjoint ${\bf 24}$ Higgs representation, so as to preserve the rank of
the group. One has then $SU(5)_{\longrightarrow}^{\langle {\bf 24}\rangle}
SU(3)\times SU(2)\times U(1)$. Since the SM fermions can acquire mass
only through the bilinears
\begin{equation}
\matrix{\bar{\bf 5} \times \bar{\bf 5}& = & \overline{\bf 10}+\overline{\bf 15}\cr
{\bf 10}\times {\bf 10} & = & \bar{\bf 5} +{\bf 45}+{\bf 50}\cr  
\bar{\bf 5} \times {\bf 10} & = & {\bf 5}+\overline{\bf 45}}
\end{equation}
it is clear that one cannot make a singlet by contracting these
fermion bilinears with the ${\bf 24}$. Hence, SM fermions do not acquire
masses in the SU(5) breaking stage (what is fortunate of course, since
they would be otherwise too heavy). To give masses to both $\bar{\bf 5}$
and ${\bf 10}$ representations with just one Higgs multiplet, one has
then to use either the ${\bf 5}$ or the ${\bf 45}$ representations. 
The minimal SU(5) model uses just one ${\bf 5}$ of Higgses $\Phi_i$ to give
fermions a mass in the process of electroweak symmetry breaking
$SM^{\langle {\bf 5}\rangle}_{\ \to} SU(3)\times U(1)_{em}$.
This Higgs representation consists of a color triplet state $T$ and
the usual Higgs doublet $H$, i.e. ${\bf 5}=(T,H)^{\tt T}$. The Yukawa
couplings can be written as
\begin{equation}
-{\cal L}_Y=f^{(1)}_{\alpha\beta}\chi_{\alpha ij}^{\tt T}C\chi_{\beta kl}\Phi_m
\varepsilon^{ijklm} +f^{(2)}_{\alpha\beta}\chi_{\alpha ij}C\Psi^i_\beta\Phi^{\dagger j}+
h.c.,
\end{equation}
where $\chi_{ij}$ are the fermions in the ${\bf 10}$ and $\Psi^i$ those in
$\bar{\bf 5}$, while $\alpha,\beta$ are family indices. When ${\bf 5}$ acquires
a VEV  breaking the electroweak
symmetry, i.e. $\langle \Phi\rangle=(0,0,0,0,v)$, 
one gets masses for the fermions, and those of 
 the down quarks and leptons satisfy
\begin{equation}
M_{\beta\alpha}^{(d)}=M_{\alpha\beta}^{(\ell)}=vf^{(2)}_{\alpha\beta}.
\end{equation}
This relation would imply in particular the equality $m_b=m_\tau$,
which again has to be checked at the GUT scale rather than at low
energies. To take into account the running of fermion masses one has to
evaluate the diagram in Figure~\ref{mrun}, which gets
contributions
from color and hypercharge gauge bosons in the loop (since $SU(2)_L$
bosons cannot provide the required chirality flip). From these one gets
\begin{equation}
{m_b(\mu)\over  m_\tau(\mu)}=\left({g_3(\mu)\over
g_{GUT}(M_X)}\right) ^{8/(11-2N_f/3)}
\left({g_1(\mu)\over
g_{GUT}(M_X)}\right) ^{3/N_f}.
\end{equation}

This gives at low energies ($\mu\simeq {\rm GeV}$) the prediction $m_b\simeq
3 m_\tau$, which is in rough agreement with the measured
values (we have here neglected the effects of Yukawa couplings).
 This seems to be then another success of the GUT idea, but
when looking however to the first two generations, the same relations
would imply $m_s\simeq 3m_\mu$ and $m_\mu/m_e\simeq m_s/m_d$, which
are however badly violated, since experimentally $m_\mu\simeq m_s$ and 
$m_\mu/m_e\simeq 10 m_s/m_d$.
One proposed solution to fix this is to include also a ${\bf 45}$
Higgs representation to generate the fermion masses \cite{geo79}. This one has the
property of introducing a Clebsch Gordan factor of --3 for the lepton
mass term with respect to that of the down quarks, so that the
prediction arising from the ${\bf 45}$ induced mass term alone would
be $m_\ell=3m_d$ at the GUT scale. Running this then down to low
energies would lead to $m_\ell\simeq m_d$, as seems to be the case
for the second generation. Hence it would seem desirable to exploit
the coupling to the ${\bf 5}$ to generate the third generation masses,
while that to the ${\bf 45}$ to generate the second generation ones.
Furthermore, with appropriate Yukawas (with `texture' zeroes imposed
by discrete symmetries) one may fix the masses of the three
generations. For instance, taking
\begin{equation}
\lambda_{\bf 5}\langle{\bf 5}\rangle\sim \pmatrix{0 & A & 0\cr A & 0 &
0\cr 0 & 0 & B}\ \ ,\ \ 
\lambda_{\bf 45}\langle{\bf 45}\rangle\sim \pmatrix{0 & 0 & 0\cr 0 & C
& 0\cr 0 & 0 & 0 }
\label{textures}
\end{equation}
one gets, after taking into account the above mentioned Clebsch, that
\begin{equation}
M_d\propto \pmatrix{0 & A & 0\cr A & C & 0\cr 0 & 0 & B}\ \ ,\ \ 
M_\ell\propto \pmatrix{0 & A & 0\cr A & -3C & 0\cr 0 & 0 & B}.
\end{equation}
This gives $m_b=m_\tau\simeq B$, which is good at the GUT scale. For
$B\gg C\gg A$ one has also $m_s\simeq C\simeq m_\mu/3$ and $m_d\simeq
A^2/C\simeq 3m_e$. When running down to low energies one obtains the
satisfactory results $m_b\simeq 3m_\tau$, $m_s\simeq m_\mu$ and 
$m_\mu/m_e\simeq 9 m_s/m_d$. Of course the `textured' Yukawas in
Eq.~(\ref{textures}) were put in by hand, but the success of the
predictions may be giving a hint on the kind of symmetries required to
generate them.

\subsection{GUTS AND PROTON DECAY}
In the SM there is no way to write renormalizable terms in the
Lagrangian   consistent with the gauge symmetries but  violating baryon
number $B$. 
Hence, one has the fact that $B$ is an accidental symmetry
 (it was not imposed by any deep fundamental reason)  of the SM
Lagrangian, and is only violated by the non-perturbative effects
related to the anomaly mentioned before. As a consequence, the proton
turns out to be stable in the SM, since it is the lightest particle with
non-zero $B$. When GUTs are considered, we have seen that quarks and
leptons reside together in large GUT multiplets. This means that
there will be gauge bosons connecting them, and hence violating $B$ and
$L$. For instance, of the 24 gauge bosons in the adjoint of $SU(5)$
there are, besides the 12 ones belonging to the SM, additional color
triplet and weak doublets vector fields $(X_\mu^\alpha,Y_\mu^\alpha)$
(with $\alpha=1,2,3$ the color index), which together with their
antiparticles make the twelve remaining gauge bosons. They have
electric charges $Q(X)=4/3$ and $Q(Y)=1/3$, and couple to the fermions
through 
\begin{equation}
{\cal L}\supset {g_5\over \sqrt{2}}\left\{X^\alpha_\mu\left[\bar
d^c_\alpha \gamma^\mu e+\bar d_\alpha \gamma^\mu
e^c+\varepsilon_{\alpha\beta\gamma}\bar u^{c\gamma}\gamma^\mu
u^\beta\right]+  
Y^\alpha_\mu\left[-\bar
d^c_\alpha \gamma^\mu \nu-\bar u_\alpha \gamma^\mu
e^c+\varepsilon_{\alpha\beta\gamma}\bar u^{c\gamma}\gamma^\mu
d^\beta\right]\right\}. 
\end{equation}

\begin{figure}
\begin{center}
\includegraphics[width=10.5cm]{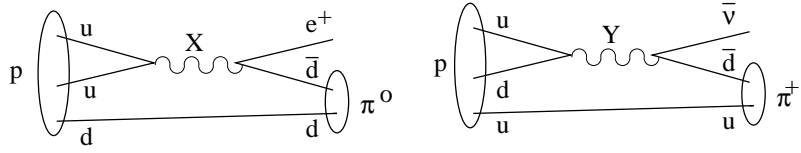} 
\caption{Diagrams inducing proton decay by gauge boson exchange in SU(5).}
\label{pdecay}
\end{center}
\end{figure}

Hence, we see that $X_\mu$ couples as a `leptoquark' ($X\to e^+\bar d$)
and as a `diquark' ($X\to uu$), clearly violating $B$ and $L$,
although always preserving $B-L=2/3$. The same happens with the
couplings of $Y_\mu$ ($Y\to \bar\nu\bar d, \ ud$). Combining these
couplings one can construct the proton decay diagrams shown in
Figure~\ref{pdecay}, which lead to $p\to e^+\pi^0,\ \bar\nu\pi^+$. 
There are similarly diagrams leading to $n\to \bar\nu\pi^0$, 
although no $n\to e^-\pi^+$ is allowed by $B-L$
conservation. When contracting the heavy gauge boson propagator one
gets an effective dimension six operator $qqq\ell/M_X^2$  which is the
one responsible for the p-decay. To compute the decay rate is somewhat
delicate, since QCD corrections with gluons exchanged between the
external quark lines are sizeable (overall factor $\sim 3$),  and there
are also significant uncertainties coming from the hadronic matrix
elements, which have to be estimated from hadronic bag models or QCD
sum rules. As a result, one gets typical estimates for the proton
lifetime
\begin{equation}
\tau(p\to e^+\pi^0)\simeq (0.2-8)\times 10^{31}\left({M_X\over
10^{15}\ {\rm GeV}}\right)^4{\rm yr}.
\end{equation}
Although this lifetime is huge, it is measurable by looking to large
quantities of protons (1~kton of water has some $10^{33}$ 
protons). Actually the search for proton decay to test GUT predictions
was the original purpose of the large underground detectors IMB and
Kamiokande. However, these searches proved to be in vain, and the
present bound set by Superkamiokande is 
$\tau(p\to e^+\pi^0)>1.6\times 10^{33}$~yr, 
clearly excluding the simplest version of  SU(5) GUT.

\subsection{THE DOUBLET--TRIPLET SPLITTING PROBLEM}
\begin{figure}
\begin{center}
\includegraphics[width=7.5cm]{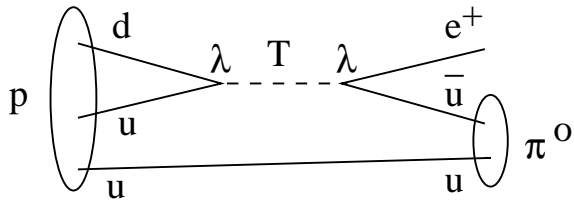} 
\caption{Diagram inducing proton decay by Higgs triplet exchange in SU(5).}
\label{tripletpd}
\end{center}
\end{figure}

We have seen that in SU(5) the SM Higgs doublet belongs to a ${\bf 5}$ 
in which also a triplet is present. This colored triplet also
mediates proton decay through the diagram in
Figure~\ref{tripletpd}. Although the diagram is suppressed by Yukawa
couplings as compared to the gauge boson mediated ones, it would
anyhow lead to extremely rapid proton decay unless the triplet states
are sufficiently heavy, typically $m_T>10^{12}$~GeV. Making the 
triplet state so heavy, while at the same time keeping the SM doublet
Higgs sufficiently light ($m_H<{\rm TeV}$)  constitutes the so-called
doublet-triplet splitting problem, which is one of the challenges for
the GUT theoreticians. 
Writing the most general scalar potential with the Higgses present in
minimal SU(5), i.e. with ${\bf 5}$ and ${\bf 24}$,  gives rise to mass
terms for both $H$ and $T$ of the order of the GUT scale.  It is
possible however to fine-tune the couplings in the potential so that
$m_H=0$ at tree level, but this relation is generally not stable  under
radiative corrections. One possible solution, the ``missing partner''
mechanism, is to couple the ${\bf 5}$ to a multiplet containing
triplet states but no doublets (such as the ${\bf 50}$ in SU(5)), and
hence through this coupling, which also involves the ${\bf 75}$, only
the triplet acquires  a (large) mass, with the
doublet remaining light.
Another possibility 
 is to start with a scalar potential with a
large global symmetry (larger than the GUT gauge one), and such that
when the symmetry is broken the doublet Higgs boson remains as a
Goldstone boson of the spontaneously broken 
global symmetry \cite{in86}, being then naturally light at tree level  without the
need of fine-tuning the couplings.  Yet another possibility (and there
are many more) which has received attention recently involves the GUT
breaking by orbifold compactification of extra dimensions \cite{ka01}.

\subsection{GUTS AND $m_\nu$} 
Lacking any observation of proton decay, nucleon decay experimenters
turned then to study just the background in their detectors, which
consists of the atmospheric neutrinos produced by the cosmic rays
hitting the top of the atmosphere. These lead to the greatest success
of those experiments through the observed deficit in the ratio of
$\nu_\mu/\nu_e$ fluxes. In particular, the zenith angle dependence of 
this ratio measured by the SuperKamiokande experiment lead to a very
clean signal supporting neutrino oscillations (such that muon
neutrinos oscillate into tau neutrinos in their way from the top
of the atmosphere up to the detectors, and at multi-GeV energies can do
that efficiently when coming from below but don't have enough time to
oscillate when coming from above). The implied neutrino mass difference
$\Delta m^2\simeq 3\times 10^{-3}$~eV$^2$ constitutes indeed the main evidence 
in favor of physics beyond the SM that we have at present.

The favored mechanism to generate naturally small neutrino masses is
the so-called see-saw mechanism \cite{ge79}. This requires the existence of right
handed neutrinos, which being SM singlets can naturally acquire a
large mass in some stage of GUT breaking. Through the combined
action of a Dirac mass term and the singlet Majorana mass term, with 
\begin{equation}
-{\cal L}_m={1\over 2}(\bar\nu_L\ \overline{(\nu_R)^c})\pmatrix{0 & m_D\cr m_D &
 M_R}\pmatrix{(\nu_L)^c\cr \nu_R}+h.c.,
\end{equation}
one gets the light mass eigenstates with masses (here we are ignoring the
family structure) 
\begin{equation}
m_\nu\simeq {m_D^2\over M_R}\simeq 10^{-3}{\rm eV}\left( {m_D\over
{\rm GeV}}\right)^2\left({10^{12}\ {\rm GeV}\over M_R}\right),
\end{equation}
so that the lightness of the neutrinos is just due to the heaviness of
the right handed states.

The necessary 
appearance of right handed neutrinos in GUTS such as SO(10) is
then most welcome, and the see-saw mechanism naturally fits within
those models (and in general in left-right GUT models also).

One difficulty however with SO(10) is that to give a mass to $\nu_R$ 
through a Yukawa coupling involving the fermion bilinear 
${\bf 16}\times {\bf 16}={\bf 10}+{\bf 126}+{\bf 120}$   requires the  
introduction of a large Higgs multiplet, the ${\bf 126}$,  since by
decomposing the SO(10) multiplets above into their SU(5) content, it
can be checked that only the ${\bf 126}$  contains the singlet state
corresponding to the bilinear right handed neutrino combination. 
Such large representations are
however not found for instance when obtaining these GUT theories
from the field theory limit of a superstring theory. 
One possibility to give mass to the right-handed
neutrino in SO(10) without introducing a ${\bf 126}$ is to do it
radiatively 
with two vacuum insertions of a scalar ${\bf 16}$ effectively acting as a ${\bf
  126}$, or more generally just using non-renormalisable couplings. 

\subsection{GUTS AND BARYOGENESIS}
Very soon after the observation of CP violation in the Kaon system, Sakharov
realized that it was possible to generate dynamically the baryon asymmetry
through microphysical processes taking place in the early Universe. Besides
the CP (and C) violation, which account for the asymmetry between particles
and antiparticles, he found that it was necessary to have baryon number
violating interactions (so as to have $B\neq 0$ at the end starting from an
initial state with $B=0$) and to be out of equilibrium, so that these
same interactions do  not erase the
generated baryon asymmetry. At the time there were no theories predicting
$B$-violating interactions, so that when GUT models were proposed to unify the
gauge interactions, the prediction that they should lead to $B$-violating
couplings was very welcome by cosmologists. Since this $B$-violation is linked
to very heavy particles (the triplet Higgses $T$ or the $X,Y$ gauge bosons for
the SU(5) case), the departure from equilibrium could take place just after
these particles become non-relativistic in the very hot Universe. The
annihilation and decay rates at this stage may not be fast enough to keep the
densities at their equilibrium values in the rapidly cooling Universe.
The traditional scenario for baryogenesis was for instance to have the heavy
SU(5) Higgs triplets decaying out of equilibrium and through $B$ and $CP$
violating decay channels. $CP$ violation implied e.g. $\Gamma(T\to
uu)\neq\Gamma(T^*\to \bar u\bar u)$, and this can occur once loop effects are
considered. 

However, when in the eighties it was realized that sphaleron mediated $B$
violation in the SM was unsuppressed at temperatures $> 100$~GeV \cite{ku85}, it was clear
that this could have the effect of erasing any asymmetry produced previously at the GUT
stage. Electroweak anomalous processes change both $B$ and $L$, but leaving intact
$B-L$. Hence, the way out was to have a GUT theory generating initially a
non-zero $B-L$, but this was not the case in the simplest SU(5) model, which
preserves $B-L$. The other possibility would be to have baryogenesis at (or
below) the electroweak scale, when sphalerons are no more active.

An interesting proposal is the so-called leptogenesis \cite{fu86}, 
in which the heavy
right-handed neutrinos of the see-saw model decay at early times violating
lepton number (with $CP$ violation resulting in $\Gamma(N\to LH)\neq \Gamma(N\to
\bar L H^*)$). The lepton asymmetry so produced is later partially transformed
into a baryon asymmetry by the sphaleron processes. This is at present the
most natural way of explaining the observed excess of matter over
antimatter, since it 
just requires the see-saw mechanism and to be not particularly unlucky with
the choice of model parameters.

\section{SUPERSYMMETRY}
\subsection{THE NATURALNESS PROBLEM}
In general one says (following t'Hooft) that a small parameter in a theory is
natural when setting it to zero increases the symmetry of the problem, so that
this very same symmetry is the responsible for the smallness of the
parameter. 
For
instance, when applied to particle masses one has that the masslessness of a
vector field can be related to the gauge invariance of the theory (as
in the case of the photon), while the
vanishing of a Dirac fermion mass is associated to a chiral symmetry (and the
vanishing of a neutrino Majorana mass may be associated to a lepton number
symmetry).  However, for a scalar field in general no symmetry is
gained when setting the mass to zero, except in the very particular 
case in which the
boson is the Goldstone boson of a global symmetry. This implies that
even if we set by hand the tree level scalar mass to zero, there is no
symmetry protecting it from acquiring large (quadratically divergent)
corrections at the loop level\footnote{Even in the case that we 
mentioned before in which the doublet Higgs boson is kept light in a
GUT theory by making  it a Goldstone boson of 
a spontaneously broken global symmetry of the
scalar potential, this  symmetry does not prevent the appearance of 
quadratically
divergent corrections to its mass from diagrams involving gauge or
Yukawa interactions.}.
These quadratic divergences make the low energy model with
fundamental scalars very sensitive to the ultraviolet structure of the
theory, and the small scalar masses are then unnatural. The solution
of this dilemma can be either to abandon the concept of fundamental
scalars, as in technicolor models, or to search for a theory where the
offending quadratic divergences miraculously cancel. 
Since fermion loops have
opposite sign as bosonic loops, a theory associating to each fermion a
bosonic partner, and relating their couplings so as to ensure the
cancellation of the quadratically divergent loop corrections
associated to them, would be able to do the miracle. This is what is 
called supersymmetry, which is a symmetry relating bosons with
fermions, and in so doing  allows in some sense for chiral symmetry to
protect also bosonic masses. An important point is that in order that
the quadratically `divergent' loop corrections to the SM Higgs boson
mass do not exceed the electroweak scale (and hence the expected value
of the Higgs
boson mass), i.e. $(g^2/16\pi^2)\Lambda^2<(250$~GeV)$^2$, 
the cutoff
signaling the energy at which supersymmetry should enter to play a
role  should be $\Lambda_{SUSY}<(4\pi/g)250$~GeV$\sim$TeV. Hence,
the solution of the naturalness problem requires that supersymmetry be
present at the weak scale, and not much above.

\subsection{SUPERSYMMETRIC MULTIPLETS}
What supersymmetry does is to pair together fermionic and
bosonic fields, and mix them up through the supersymmetric
transformations, which can schematically be written as
\begin{equation}
\delta_{SUSY}{\rm Fermion}={\rm Boson}\ \ ,\ \ \delta_{SUSY}{\rm Boson}= 
{\rm Fermion}.
\label{susy}
\end{equation}
On the other hand  two supersymmetric transformations can produce a
translation since  
\begin{equation}
\{\delta_{SUSY},\delta_{SUSY}\}={\rm translation},
\end{equation}
and hence supersymmetry is intimately related to Poincar\'e
invariance. 
There is indeed a theorem (due to Haag, Lopusza\'nski and
Sohnius) stating that the most general symmetry of the $S$ matrix is
just SUSY$+$Poincar\'e$+$internal symmetries. Moreover, the fact that
supersymmetric transformations can generate translations indicates that
making supersymmetry a gauge symmetry (so that the transformation
parameters depend on the space-time point) should lead to invariance under
general coordinate transformations and hence  automatically
include General Relativity. Local supersymmetry (supergravity) may
then be the road to incorporate
gravity into the SM.
 
Since SUSY transforms fermions into bosons and vice-versa, it requires
the existence of fermionic generators $Q_\alpha$, and the above
mentioned relation between two SUSY transformations and a translation
results from the SUSY algebra relation
$\{Q_\alpha,\bar Q_{\dot\beta}\}=2\sigma^\mu_{\alpha\dot\beta}P_\mu$
(with the dot over the index just indicating that it refers to the
conjugate of the generator, and $\sigma^\mu=(1,\vec\sigma)$). A
useful concept to write down supersymmetric  lagrangians is that of
superspace, introduced by Salam and Strathdee \cite{sa70}. 
The superspace includes a
 fermionic `coordinate' $\theta_\alpha$ which is somehow  conjugate to
the 
supersymmetric generator $Q_\alpha$, in the same sense that the usual 
coordinates
$x_\mu$ are conjugate variables to $P_\mu$. The supersymmetric
particle multiplets are then described by superfields defined in the
superspace $(x_\mu,\theta_\alpha)$.

The basic multiplets of a supersymmetric theory are:
\begin{itemize}
\item The chiral supermultiplet
$\Phi(x,\theta)$ contains a
complex scalar $A$, a Weyl fermion $\Psi$ and an auxiliary scalar $F$
(this last is not propagating since its equation of motion is just
algebraic, involving no derivatives, and hence it can be eliminated in
favor of the other two propagating fields). Under an infinitesimal global
supersymmetric transformation with parameter $\xi$ one has, as anticipated in
Eq.~(\ref{susy}),  that
$\delta A=\sqrt{2}\xi\Psi$  and $\delta\Psi=\sqrt{2}(\xi
F+i\sigma^\mu\bar\xi\partial_\mu A)$. 
On the other hand, the transformation of the 
auxiliary field is  a total derivative, $\delta
F=\sqrt{2}i\partial_\mu\bar\Psi\sigma^\mu\bar\xi$. 
\item The real vector multiplet $V^a$ is used to describe gauge fields
$A_\mu^a$, and
has the associated fermionic partners $\lambda^a$ known as gauginos, which will 
then be in the adjoint representation of the gauge group. The
superfield associated to them has also an auxiliary component, the $D^a$
scalar field, which also transforms into a total derivative under an
infinitesimal SUSY transformation. 
\item The graviton supermultiplet involves the spin two graviton
$G_{\mu\nu}$ paired to its superpartner, the spin $3/2$ gravitino.
\end{itemize}

The trick to write down a supersymmetric Lagrangian is to exploit the
fact that auxiliary $F$ terms of chiral multiplets and $D$ terms of
vector multiplets transform under SUSY as total derivatives, and are
hence good Lagrangian densities. For instance, one can construct a
vector superfield from the following combination of the matter chiral
superfields $\Phi_i$ and gauge boson 
 vector superfield $V^a$, 
\begin{equation}
\sum_{i,j}\Phi_i^\dagger {\rm exp}[2gT^a_{ij}V^a]\Phi_j.
\end{equation}
It can be seen that the $D$ term (auxiliary part) of this superfield, 
when written in terms of component fields,  includes the
 usual fermion covariant derivative involving the coupling to the 
gauge bosons, and is hence the appropriate supersymmetric
generalization of the gauge invariant fermion kinetic term. This
Lagrangian density also includes new terms such as a fermion-sfermion-gaugino
coupling (with the same strength as the previous one as required by
 supersymmetry). Furthermore, it also includes a
quartic scalar coupling $\sum_a(D^a)^2/2$, where
\begin{equation}
D^a=g\sum_i\Phi^\dagger_iT^a\Phi_i.
\end{equation}
The strength of this quartic scalar coupling is $g^2$, as 
 required in order that the quadratic
divergences in the sfermion self energies coming from the scalar loop
associated to this quartic coupling (summed to the loop involving the gauge
bosons)  cancels with the one coming from
the fermion-gaugino loop. 

The gauge kinetic terms result from the $F$ component of the chiral
field obtained from the square of the so-called field-strength chiral
field,
${\cal W}=\lambda+(D+i\sigma^{\mu\nu}F_{\mu\nu}/2)\theta/2+
\theta^2\sigma^\mu\partial_\mu\bar\lambda/4$. 

Finally, the Yukawa
couplings result from the $F$ term of the superpotential $W$, which is a
generic cubic (so as to be renormalizable) polynomial in the chiral
fields, i.e.
\begin{equation}
W(\Phi)={1\over 2}m_{ij}\Phi_i\Phi_j+{1\over
3}\lambda_{ijk}\Phi_i\Phi_j\Phi_k.
\end{equation}
In order for $W$ to be a chiral superfield, it is necessary that it be
an holomorphic function of the chiral superfields $\Phi_i$, not
involving the conjugate fields, and this has important
phenomenological implications.
For instance, in the minimal supersymmetric standard model (MSSM) two
Higgs doublets of opposite hypercharge are required to give masses to
up-type and down-type quarks. This is different than what happens in
the SM, where one can
freely use the conjugate Higgs doublet $\sigma_2H^*$ (which has
opposite hypercharge as $H$) together with $H$ for this purpose,
but this is not derivable from a superpotential in a supersymmetric
framework. Another reason justifying the need for two Higgs doublets
is that the higgsino fermionic partners contribute to triangle
anomalies, and hence two higgsinos with opposite charges are required
to cancel them.

Besides giving rise to the Yukawa couplings ${\cal
L}_Y=(1/2)(\partial^2W/\partial\Phi_i\partial\Phi_j)\Psi^i\Psi^j$, the
superpotential also generates a contribution to the scalar potential
depending on $F_i\equiv \partial W/\partial \Phi_i$. The complete
scalar potential is then 
\begin{equation}
V=\sum_i|F_i|^2+{1\over 2}\sum_a(D^a)^2.
\label{potential}
\end{equation}
Again the quadratic divergences of the scalar fermion masses
associated to the quartic couplings present in $|F|^2$, which are
proportional to $\lambda^2$, cancel with those from the ordinary
fermion loops induced by the Yukawa couplings $\lambda$.

\subsection{SUPERSYMMETRY BREAKING}
Following the procedure described in the previous section one can then
write down a supersymmetric Lagrangian involving the standard model
fields and their superpartners (the scalar fermions: sleptons,
sneutrinos and squarks; the gauge fermions: gluinos, winos and binos;
the Higgs fermions or higgsinos). Clearly an immediate drawback of
this would be the appearance of all these states with the same charges
and masses as the SM ones but different spin, which are not observed
in nature. This implies that supersymmetry has to be broken, so as to
lift the degeneracy in mass inside supermultiplets and make the
superpartners sufficiently heavy and beyond present experimental
bounds.
  However, to retain
the good properties of SUSY, this breaking has to be `soft',
i.e. without reintroducing quadratic divergences \cite{gi82}.  This means that
only some kind of SUSY violating terms in the Lagrangian are allowed,
such as bilinear or trilinear scalar couplings (but not quartic ones)
and gaugino masses (but not fermion Yukawa couplings without their
associated $|F|^2$ terms in the potential). 
The SUSY soft breaking Lagrangian can then be
written as
\begin{equation}
-{\cal
 L}_{soft}=\sum_{scalar}m_i^2|\phi_i|^2+
(AW^{(3)}+BW^{(2)}+\sum_aM_a\lambda_a\lambda_a+h.c.) ,
\end{equation}
where $M_a$ are gaugino masses, while 
\begin{equation}
W^{(3)}=\lambda_uH_2\tilde Q\tilde u^c+\lambda_dH_1\tilde Q\tilde
d^c+\lambda_\ell H_1\tilde L\tilde \ell^c
\end{equation}
contains the cubic terms in the superpotential, but with the
superfields replaced by their scalar components. Similarly, the
quadratic term is $W^{(2)}=-\mu H_1H_2$.

We have not specified the family structure of the different couplings,
but once this one is taken into account, it is found that the most
general softly broken supersymmetric Lagrangian contains more than one
hundred new
unspecified parameters, clearly introducing a lot of arbitrariness in the
model building. Moreover, generic parameters can induce large FCNC
effects and violate bounds on $\Delta m_K/m_K$, on $K_L\to
\mu^+\mu^-$ or $\mu\to e\gamma$, they can lead to large $CP$ violating
electric dipole moments for the neutron or the electron, they can
affect $\epsilon_K$ or $\epsilon'_K/\epsilon_K$, etc.. In particular,
the absence of a generalized GIM mechanism to suppress FCNC can be
understood from the fact that the 
fermion-fermion-neutral gauge boson coupling remains diagonal after 
a unitary rotation of the
fermion fields, and the sfermion-sfermion-gauge boson also after a
unitary rotation of the sfermion fields, but however the
fermion-sfermion-gaugino vertex will not be diagonal in the flavor
indices after going to the mass eigenstate
basis\footnote{Alternatively, one may diagonalize the
fermion-sfermion-neutral gaugino vertices and then be constrained to
work with off-diagonal sfermion mass matrices.}. 

The way to break spontaneously supersymmetry without breaking Lorentz
invariance is to induce a non-vanishing VEV for the auxiliary component
of a superfield, i.e. either $\langle F\rangle \neq 0$ ($F$ breaking)
or  $\langle D\rangle \neq 0$ ($D$ breaking). In this case the ground
state energy will be non-vanishing (see Eq.~(\ref{potential})), what
is the signal of global supersymmetry breaking.

\begin{figure}
\begin{center}
\includegraphics[width=12.5cm]{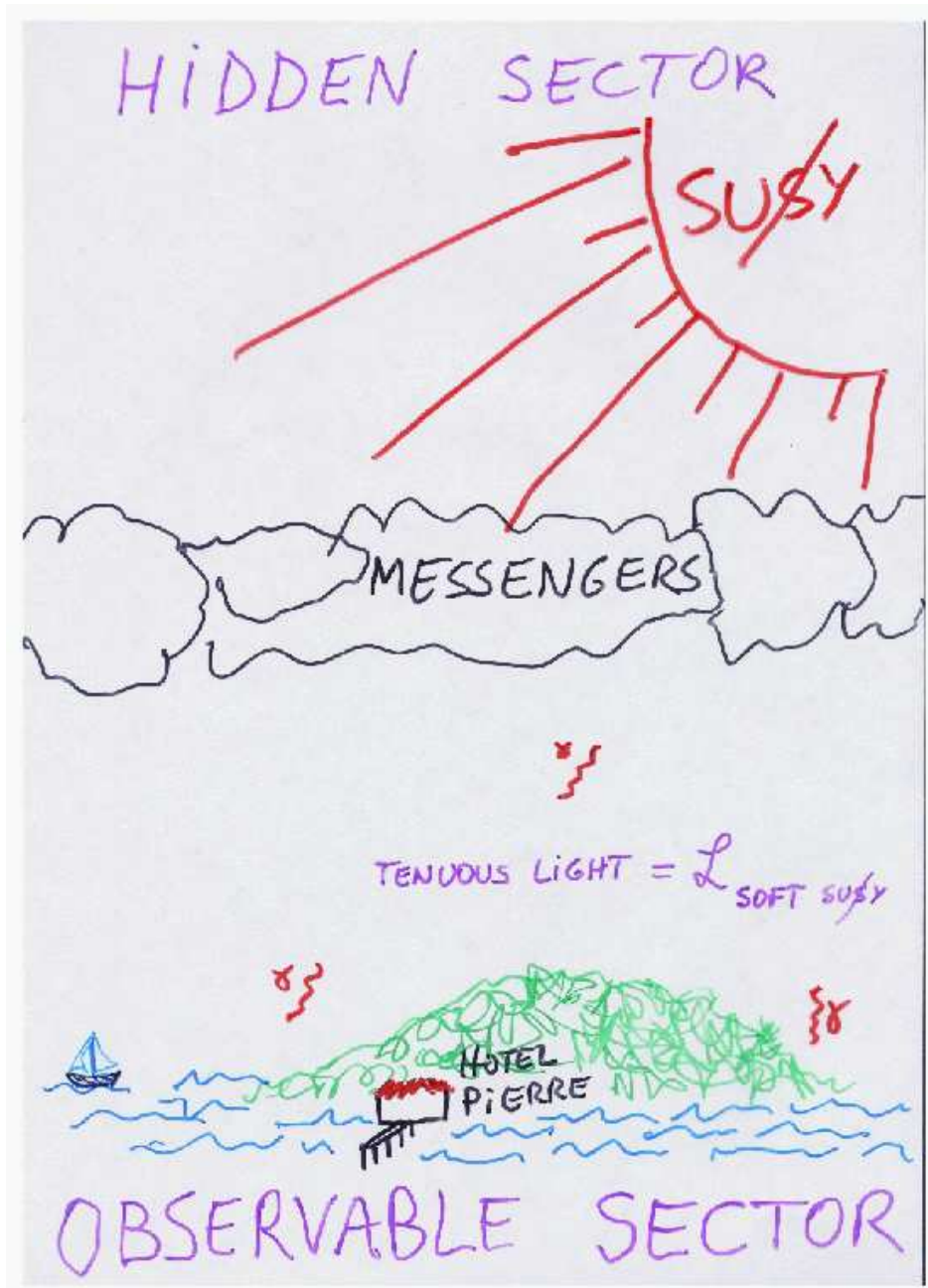} 
\caption{SUSY breaking in hidden sector and its mediation to the
observable sector.}
\label{itacu.fig}
\end{center}
\end{figure}

 In global supersymmetry
breaking there is an important constraint on the resulting mass
spectrum, which is known as the tree-level supertrace formula, i.e.
\begin{equation}
{\rm STr}{\cal M}^2\equiv \sum_{bos}{\cal M}^2-\sum_{ferm}{\cal M}^2=0.
\end{equation}
For instance, considering just a chiral multiplet with a scalar
component $\varphi=S+iP$, the typical spectrum after supersymmetry
breaking will be to have the scalar and pseudoscalar components
splitted with respect to the fermionic particle in such a way that
$M^2_{S,P}=M_F^2\pm \Delta M^2$, i.e. $M_S^2+M_P^2-2M_F^2=0$ (the two
in the fermionic contribution to the supertrace comes from the two
d.o.f. of a Weyl fermion).
This clearly cannot be implemented in the observable sector of the
supersymmetric SM, since it would imply for instance the existence of
scalar fermions lighter than the corresponding fermions (or with
negative squared masses). Hence, the usual strategy is to break
supersymmetry in a hidden sector not directly coupled to the SM
fields, 
and communicate it to the observable sector by gravitational effects
(gravity mediation), at the loop level by means of gauge interactions 
(gauge mediation) or for
instance exploiting the superconformal anomaly (anomaly mediation). 
A cartoon of this general framework is shown in Figure~\ref{itacu.fig},
which was drawn after a week of rain during the meeting 
at the Itacuru\c ca Island: the
 hidden  sector is represented by the sky above the clouds, which was
totally disconnected from the observable sector. The thick layer of
clouds represent the messenger fields, which are in contact with both
sectors and once in a while let some
photons go through, allowing the breaking of
susy (the sunshine) to be transmitted into the observable sector,
giving rise to the tenuous light that we could see.

The simplest way to suppress FCNC effects is by having a mechanism of
supersymmetry breaking mediation giving rise to universal scalar
masses $\tilde m^2$, so that in
first approximation the mass eigenstates can be freely rotated. In
gravity mediation scenarios, this is usually attributed to the
universal character of gravitational interactions, that should then
generate a common soft mass for all scalar fermions (also generation
independent $A$ and $B$ soft terms are predicted). In gauge
mediation scenarios, the scalar masses arise at two-loops and turn out
to be proportional to the gauge coupling constant ($m_{\tilde
f}^2\propto \alpha_i^2$), so that for instance squarks are degenerate
among themselves as a result of the blindness of gauge interactions to the
family indices.

It has to be noticed however that the universal conditions will hold
at the messenger scale (i.e. at the Planck scale for gravity
mediation, or at a smaller scale $M>10^2$~TeV for gauge mediation),
but these masses run for decreasing scales and in so doing induce
non-universalities in the mass spectrum. One then generally
predicts anyway the presence of
 non-zero FCNC effects at low energies, which in gravity
mediation schemes are sometimes even at the verge of contradicting
experimental bounds. Regarding the gaugino masses, one often assumes
for simplicity that there is a unification relation, so that they
become all equal at the GUT scale ($M_3=M_2=M_1$) in gravity mediation
scenarios, while in gauge mediation ones they arise at one loop and
are hence proportional to the corresponding gauge couplings. 

We see then that the new parameters of a supersymmetric model can be
significantly restricted in specific models. For instance, in the
minimal supersymmetric standard model (MSSM) with gravity mediation 
one has just the five 
parameters $\tilde m^2,\ M_\lambda,\ A,\ B,$ and $\mu$.

 A lot of work has been devoted  
to try to understand the origin of soft terms and to 
relate the low energy predictions with  the possible
mechanisms of supersymmetry breaking
and mediation, which are the biggest unknowns for supersymmetry model
builders. 

\subsection{THE SUPERSYMMETRIC HIGGS SECTOR}
As we already mentioned, the supersymmetric SM requires the existence 
of at least
two complex Higgs doublets with opposite hypercharges
\begin{equation}
H_1=\pmatrix{H^0_1\cr H^-_1}\ \ ,\ \ H_2=\pmatrix{H_2^+\cr H_2^0}.
\end{equation}
This represents 8 d.o.f., and when the electroweak symmetry gets
broken 3 will be eaten by the massive gauge bosons $W^\pm$ and $Z$,
leaving one physical charged Higgs $H^\pm$ and three neutral ones (one
pseudoscalar $A$ and two scalars, $h$ and $H$).

The neutral Higgs potential will have the following contributions:
\begin{itemize}
\item The soft mass terms: $m_{H_1}^2|H^0_1|^2+m_{H_2}^2|H^0_2|^2$
\item The bilinear soft term: $BW^{(2)}=-B\mu H^0_1H^0_2$
\item The $|F|^2$ terms
\begin{equation}
\sum_i|F_i|^2\supset \left| {\partial W^{(2)}\over \partial
H^0_1}\right|^2+
\left| {\partial W^{(2)}\over \partial H^0_2}\right|^2=\mu^2\left(
|H^0_1|^2+|H^0_2|^2 \right).
\end{equation}
\item The $D$ terms
\begin{equation}
{1\over 2}D^2={1\over 2}\left(g^2\left(
T_3(H^0_1)|H^0_1|^2+T_3(H^0_2)|H^0_2|^2\right)^2+g'^2\left(
Y(H^0_1)|H^0_1|^2+Y(H^0_2)|H^0_2|^2\right)^2\right).
\end{equation}
\end{itemize}
This leads to the scalar potential
\begin{equation}
V(H^0_1,H^0_2)=m_1^2|H^0_1|^2+m_2^2|H^0_2|^2-m_3^2(H^0_1H^0_2+h.c.)+{g^2+g'^2\over
8}\left(|H^0_1|^2-|H^0_2|^2\right)^2,
\end{equation}
with $m_{1,2}^2\equiv m_{H_{1,2}}^2+\mu^2$ and $m_3^2\equiv
B\mu$. Notice that the direction $|H^0_1|=|H^0_2|$ is `$D$ flat', and
in order that the potential be bounded from below along it one needs
to satisfy $m_1^2+m_2^2>m_3^2$. On the other hand, the origin is
unstable (and hence the electroweak symmetry is broken) as long as
$m_1^2m_2^2<m_3^4$. At the GUT scale these two conditions are
incompatible under the assumption of universality, i.e. if
$m_1^2=m_2^2=\tilde m^2+\mu^2$, but a quite remarkable property of the
MSSM is that in the running to low energies, the effects of the large
top Yukawa coupling pushes down the parameter $m_2^2$, which can
become negative leading to what is known as the radiative breaking of
the electroweak symmetry. Actually this desireable feature provided
the first theoretical hint for the need to have a large top mass
\cite{ib83},  at a
time when most phenomenologist believed that the top mass had to be at the few
tens of GeV level.

The minimization of the Higgs potential leads to a vacuum state with 
\begin{equation}
\langle H^0_1\rangle={v\over \sqrt{2}}\cos\beta\ \ ,\ \ 
\langle H^0_2\rangle={v\over \sqrt{2}}\sin\beta,
\end{equation}
with 
\begin{equation}
\sin 2\beta={2m_3^2\over m_1^2+m_2^2}
\label{sbeta}
\end{equation}
and in order to reproduce the correct electroweak scale one needs to
satisfy 
\begin{equation}
{M_Z^2\over 2}={g^2+g'^2\over 8}v^2={m_1^2-m_2^2\tan^2\beta\over \tan^2\beta-1}.
\label{mz}
\end{equation}
Equation (\ref{sbeta}) is usually employed to trade the parameter $B$
by $\tan\beta\equiv \langle H^0_2\rangle/\langle H^0_1\rangle$, while
the constraint in Eq.~(\ref{mz}) is usually employed to express
$\mu^2$ in terms of the remaining susy parameters (leaving sign($\mu$)
undetermined). Hence, after imposing the electroweak symmetry breaking
constraints the additional parameters present in the gravity mediated 
MSSM, assuming universality and unification, are
\begin{equation}
\tilde m^2,\ M_\lambda,\ A,\ \tan\beta,\ {\rm sign}(\mu).
\end{equation}

The mass spectrum in the Higgs sector is directly obtained by
expanding the scalar potential around its minimum, and in so doing
one finds that the lightest Higgs boson mass satisfies $m_h<M_Z|\cos
2\beta|$. This tree-level relation (which is essentially excluded by
present LEP bounds)  is sizeably affected by one-loop
corrections to the scalar potential arising from the large top Yukawa
coupling. These lead to a contribution $\delta m_h^2\simeq
(3/\pi^2)(m_t^4/v^2) {\rm log}(\tilde m/v)$, which can bring the Higgs
mass above present bounds, but anyhow a generic prediction of the MSSM
is that the lightest Higgs should be lighter than $\sim
130$~GeV. Hence, the experimental search for a Higgs boson in this mass
range is a very important test for supersymmetry, since failing to
find it would exclude the most natural models of weak scale
supersymmetry. 

Another important aspect of the radiative corrections to the Higgs
potential is that they reduce its overall scale-dependence, 
since as we
mentioned before the tree level parameters $m_i^2$ where running
significantly (some even changing sign) as the weak scale was
approached, and hence the Higgs spectrum and couplings obtained from
the tree level potential would have a strong scale dependence, but
this is cured by the radiative corrections.

The Higgs boson searches at LEP have focused mainly on the
Higgs-strahlung process ($e^+e^-\to Z\to Zh$), from which essentially
all the kinematically allowed range $m_h<\sqrt{s}-M_Z\simeq 113$~GeV
has been excluded (for $\tan\beta<8$, since otherwise the $ZZh$
coupling is suppressed). For large tan$\beta$ the preferred discovery
channel is $Z\to hA$, i.e. the light Higgs production in association
with the pseudoscalar $A$, and this has  excluded the range  $m_h<90$~GeV.

\subsection{THE SUPERSYMMETRIC PARTICLE SPECTRUM}
\begin{figure}
\begin{center}
\includegraphics[width=8.5cm]{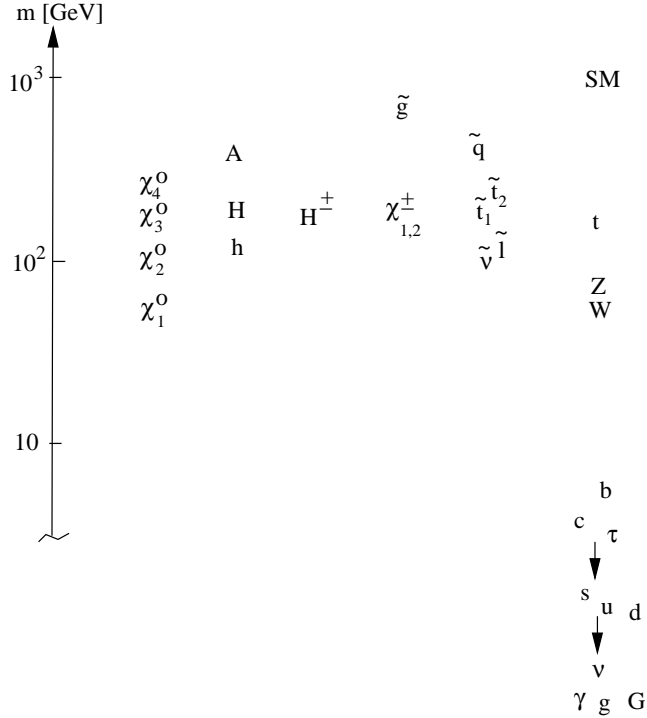} 
\caption{Plausible particle spectrum in a gravity mediated
supersymmetric model.}
\label{spectrum}
\end{center}
\end{figure}

To obtain the sparticle spectrum one has to run all soft breaking
parameters from high energies (assuming some proper boundary
conditions, probably motivated by the absence of FCNC, simplicity and
predictability) down to the weak scale. A plausible spectrum arising
from this exercise for gravity mediated scenarios 
is depicted in Figure~\ref{spectrum}. The lightest superpartner is the
lightest of the four Majorana neutralinos 
(i.e. the mass eigenstates which are
mixtures of the four neutral fermions which are the partners of the
photon, the $Z$ and the two neutral complex Higgses). The charged
fermions which are partners of the charged Higgses and of the $W$ mix
into two charginos, which are Dirac fermions since they are
charged. Colored particles get splitted from uncolored ones 
due to the running associated to strong interactions, and in this way
gluinos are typically much heavier than charginos or neutralinos, and
squarks are much heavier than sleptons. Third generation squarks are
lighter than those of the first two generations due to the effects of
the Yukawa couplings. 

Since this doubling of degrees of freedom associated to the
superpartners is expected to appear below the TeV scale, upcoming
colliders (Tevatron run II, LHC, $e^+e^-$ linear colliders) are ideal 
to search for
these particles, which indeed constitute one of their main targets. 

\subsection{SUSY AND GAUGE UNIFICATION}
\begin{figure}
\begin{center}
\includegraphics[width=10.5cm]{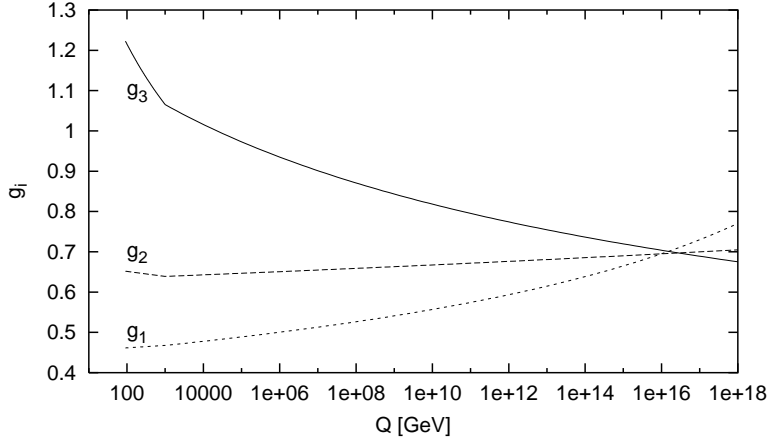} 
\caption{Running of the gauge couplings in the SUSY SM at one loop,
adopting $T_S=1$~TeV.}
\label{susyrge}
\end{center}
\end{figure}

One of the most appealing hints in favor of supersymmetry 
at the weak scale is related to the effects that it has 
on the running of the gauge
couplings. In a supersymmetric scenario, the experimentally observed 
gauge couplings do indeed unify at a unique scale, and this one is large
enough to be consistent with proton decay constraints (see
Figure~\ref{susyrge}).  

This can be seen from the change in the $\beta$ functions arising from
the additional contributions produced by loops involving gauginos and
sfermions, which for instance make SU(3) somewhat less asymptotically
free ($b_3=-3$ instead of the non-susy value of $-7$), and make the
SU(2) beta function positive ($b_2=1$), while for $U(1)_Y$ one has
$b_1=33/5$, as can be easily checked using Eq.~(\ref{beta}). These
changes delay the unification of the couplings up to $M_{GUT}\simeq
10^{16}$~GeV, and work provided the threshold scale $T_S$ above which
the supersymmetric spectrum appears is light enough,
$T_S<{\rm TeV}$.

\subsection{SUSY AND PROTON DECAY}
\begin{figure}
\begin{center}
\includegraphics[width=10.5cm]{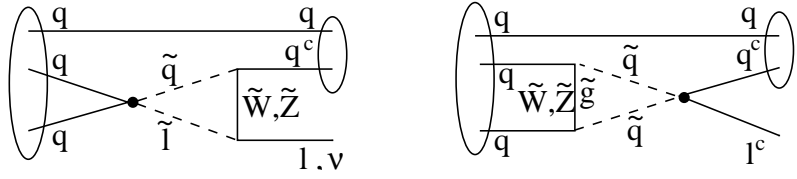} 
\caption{Supersymmetric contribution to proton decay from $dim=5$ operators.}
\label{susypdec}
\end{center}
\end{figure}

We saw that in non-supersymmetric GUTs proton decay was due to four
fermion operators of the form $qqql$ which resulted from the
contraction of the propagator of the very heavy  triplet Higgs or gauge 
bosons. Being these operators of dimension $dim=6$, they were
suppressed by two powers of the GUT scale (i.e. $\propto
M_{GUT}^{-2}$).
These lead to $\tau(p\to e^+\pi^0)\sim
10^{31}$yr$(M_{GUT}/10^{15}$GeV$)^4$, which was in contradiction with
observations for the non-supersymmetric GUT scale of $10^{14}$~GeV.
In SUSY, the GUT scale becomes $10^{16}$~GeV, and hence there is no
conflict with proton decay arising from $dim=6$ operators. In
the supersymmetric case there are however new operators, involving two bosons
(squarks or sleptons) and two fermions, which can induce proton decay,
and since these are $dim=5$, they are suppressed by only one power of
the GUT scale (i.e. by the contraction of one heavy fermionic
propagator, such as the one from higgsino triplet exchange). 
Hence, $dim=5$ operators turn out to be the dangerous
ones. Since initial and final states involve ordinary quark and
leptons, the $dim=5$ operators have to be `dressed' with light susy
partners, as is shown in Figure~\ref{susypdec}. This typically results
in the proton decay rate being proportional to $(m_{\tilde
g}\tan\beta/\tilde m^2)^2$, and hence it is quite sensitive to the
supersymmetric spectrum (increasing for heavier gluinos!). 

Another important fact
is that writing  the operators in terms of superfields, which obey Bose
statistics, one can realise the need to antisymmetrize them  with
respect to the family indices. For instance, the operator
$O_1=\epsilon^{ijk}(u^\alpha_id^\beta_j-d_j^\alpha
u_i^\beta)(u_k^\gamma\ell-d_k^\gamma\nu)$ is acceptable since it is
antisymmetric in both color ($i,j,k$) and flavor
($\alpha,\beta,\gamma$) indices. This implies that the dominant proton
decay channel in supersymmetric models is not into pions (involving
first generation quarks) but into kaons. Typical predictions for the
lifetime of this channel are
\begin{equation}
\tau(p\to K\bar\nu)\sim 10^{29}\div 10^{35}\ {\rm yrs}.
\end{equation}
The Superkamiokande bound of $\tau(p\to K\bar\nu)>6.7\times 10^{32}$~yrs
then excludes part of the supersymmetric parameter space, disfavoring 
in particular heavy gluinos for large $\tan\beta$, with the bounds
depending on the assumed scale for the scalar masses $\tilde m$.

\subsection{R PARITY SYMMETRY, COLLIDER SEARCHES AND DARK MATTER}
One important difficulty that has to be faced when trying to write
down a supersymmetric version of the standard model is that there are
Yukawa-like couplings which are consistent with gauge and super
symmetries but violate baryon or lepton number. The allowed trilinear
couplings  are
\begin{equation}
{\cal L}_{\not
R}=\lambda_{ijk}Q_iL_jU^c_k+\lambda'_{ijk}U^c_iD^c_jD^c_k+
\lambda''_{ijk}L_iL_jE^c_k.  
\label{rparity}
\end{equation}
Hence, unlike what happens in the SM, in its supersymmetric version
$B$ and $L$ are not automatic symmetries of the Lagrangian. If the
above mentioned couplings are allowed, they can give rise to extremely
fast proton decay, mediated by the exchange of squarks or sleptons
with weak scale masses, so that very strong constraints result on the
product of some pairs of couplings
(typically $\lambda\cdot\lambda<10^{-27}$!). To make these couplings so
small is unnatural, so that the simplest solution is to eliminate them
directly by invoking a symmetry, the so-called R-parity. R-parity
distinguishes between standard model particles, which are R-even
(e.g. $f\leftrightarrow f$ and $H\leftrightarrow H$ under R parity),
 and their
superpartners, which are R-odd (e.g. $\tilde h\leftrightarrow -\tilde
h$ or $\tilde f\leftrightarrow -\tilde f$). Hence, imposing R symmetry
the couplings in Eq.~(\ref{rparity}) are forbidden, while those of the
standard Yukawa couplings involving the Higgs doublet are allowed. 

The discrete R-parity symmetry 
need not necessarily be imposed by hand, and may be the left
over, after susy breaking, 
 of a continuous R symmetry (i.e. a symmetry of the susy
Lagrangian which also transforms the Grassman variables $\theta$) or it
may be related to an underlying gauge symmetry. For instance, in SO(10)
models the allowed Yukawa couplings involve the ${\bf 16}\times {\bf
16}\times {\bf 10}$, and this does not include the $R$ violating couplings
(that would be in the ${\bf 16}\times {\bf
16}\times {\bf 16}$, which is not SO(10) invariant). 

The conservation of R-parity has some very important phenomenological
implications:
\begin{itemize}
\item  It requires that superpartners be always produced in pairs, and
hence this increases the threshold for their production.
\item It implies that in the decay of a superpartner there should be
always a superpartner (or an odd number of them). This requires in
particular that the lightest superpartner (LSP) has to be stable. The
stability of the LSP, which in supergravity models is usually the lightest
neutralino and in gauge
mediation models  can be the gravitino, 
is crucial for the experimental searches, since it leads
to the characteristic missing energy signature, associated to the
escape from the detector of the weakly interacting neutral LSP. 
For instance, typical signatures of supersymmetry at $e^+e^-$
colliders involve the production of pairs of charginos $\chi^+\chi^-$
or sleptons $\tilde\ell^+\tilde\ell^-$, with their subsequent decay
into leptons or jets plus neutralinos. The first ones are observed
but the neutralinos (and eventually some neutrinos) are not, and their
effect is to take away significant amounts of energy, which will be
missing in the overall budget. Strongly interacting particles such as
squarks and gluinos are best searched at hadronic colliders, through
the processes $qq,qg,gg\to\tilde q\tilde q,\tilde q\tilde g,\tilde
g\tilde g,\tilde q\chi,\tilde g \chi$, with subsequent cascade decays
of the superpartners. Another `background free' process is the
trilepton signal with missing energy associated to the
chargino/neutralino pair production, i.e. $p\bar p\to
\chi^\pm\chi^0_2$, with $\chi^\pm\to\ell^\pm\nu\chi^0_1$ and
$\chi^0_2\to\chi^0_1\ell^+\ell^-$. 

Up to the present no signal of superpartner production has been
observed, so that roughly speaking charginos and sleptons have to be
heavier than $\sim 100$~GeV, while squarks and gluinos must be above
$\sim 200\div 300$~GeV. A major improvement in the mass reach will be
achieved with the LHC, allowing to test superpartner masses up to the
TeV scale, so that it may not be unreasonable to say that
supersymmetry at the  weak scale (as required by the naturalness
problem) will have to be discovered (or discarded) by this machine.

\item Another major implication of the LSP stability is its possible
relevance to explain the dark matter which is known to be present in
the Universe. Within supersymmetric
models, the dark matter can naturally be attributed to LSPs 
which are left over from the early stages of the hot
Universe. At that time all particle species where initially 
in thermal equilibrium and
the weakly interacting ones were `frozen out' little after they became
non-relativistic and in this way they were able to 
 survive up to the present times.  
The exact amount of relic particles that remains depends essentially
on the rate of annihilations of neutralinos since this determines how the
decoupling takes place. Larger annihilation rates clearly imply less
surviving particles, and for a weakly interacting massive particle
(WIMP, the neutralino being the preferred candidate), one can show
that the contribution to the relic density is $\Omega\simeq
10^{-37}{\rm cm}^2/\langle \sigma_{ann}v\rangle$, but of course several
subtleties enter into a detailed computation (many possible channels, 
threshold effects, resonant
annihilations, coannihilations among different superpartners, etc.).

At any rate,  it is remarkable that weak scale supersymmetry, which has its
motivations in theoretical issues completely unrelated with the dark
matter problem, predicts the existence (at least in R parity
conserving scenarios) of a natural candidate for cold dark matter. 
On the other hand, if neutralinos indeed constitute the galactic dark
matter, this  offers a new possibility to experimentally search for
supersymmetry trying to detect the WIMPs and to characterize their
properties.  Several groups are trying to observe directly the nuclear
recoils resulting from halo WIMPs interacting inside low background
detectors (using germanium, sodium iodine, superconducting granules,
liquid xenon, etc.). Others (such as Superkamiokande or MACRO) 
are trying to observe the high energy
neutrinos produced in the annihilations of the dark matter neutralinos
which have become trapped by the Sun or the Earth in the last five
Gyrs.
The predicted rates depend strongly on the supersymmetric parameters,
but for instance may be in the range of the sensitivity of upcoming
direct detection 
experiments for large values of tan$\beta$, especially for positive
$\mu$, in which case the spin independent coherent contribution to the
scattering cross sections is enhanced. There is even a claim of a
positive signal having been observed by the DAMA Collaboration, but
this is partly in
conflict with results from other groups.

\subsection{SUPERSYMMETRY IN VIRTUAL PROCESSES}

\begin{figure}[t]
\begin{center}
\includegraphics[width=10.5cm]{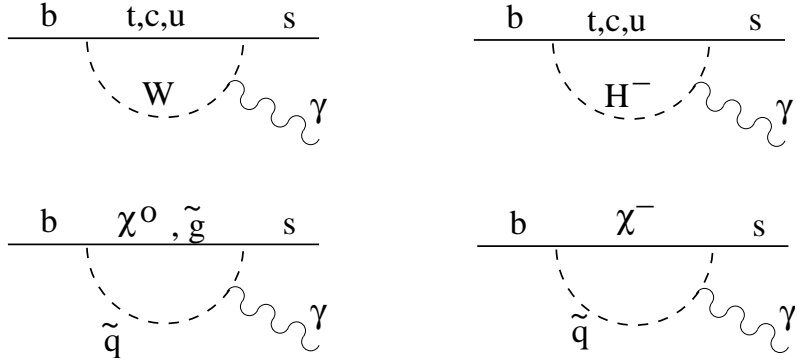} 
\caption{SM and supersymmetric contributions to $b\to s\gamma$. The photon
can be attached to any charged particle line.}
\label{figbsgam}
\end{center}
\end{figure}

Besides producing superpartners at colliders or trying to detect those
produced in the big bang, supersymmetry may manifest through the
effects that virtual superpartners may have on some particular
processes. Two important cases are the flavor changing transition
$b\to s\gamma$ and the anomalous magnetic moment of the muon.
The first one  takes place already at the loop level in the SM,
involving in particular the exchange of the heavy top quark, and the
supersymmetric contribution can naturally be of the same order of
magnitude. The diagrams contributing to it are shown in
Figure~\ref{figbsgam}. The loop involving the charged Higgs has the same
sign as the SM contribution, the one involving neutralino or gluino
exchange is usually negligible while the one involving chargino
exchange grows with tan$\beta$, and hence is the leading one for large
tan$\beta$. It interferes destructively with the SM piece for $\mu>0$
and constructively for $\mu<0$. Since the measured rates are already
somewhat below the SM expectations (see Eq.~(\ref{bsgam})), one may say that
large tan$\beta$ with negative $\mu$ is in conflict with observations,
and that positive $\mu$ is somewhat preferred.

\begin{figure}
\begin{center}
\includegraphics[width=10.5cm]{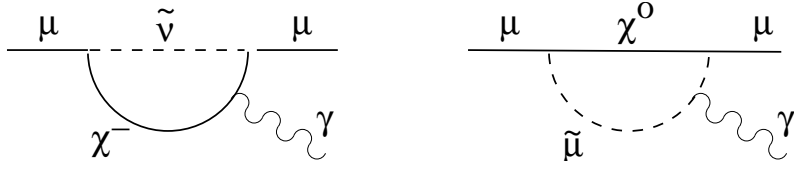} 
\caption{Supersymmetric contributions to the muon anomalous magnetic moment.}
\label{susyamu}
\end{center}
\end{figure}

Regarding the anomalous magnetic moment of the muon, the
supersymmetric contribution arises from the diagrams in
Figure~\ref{susyamu}. Since these are flavor and CP conserving
processes they are hard to avoid, and they are of a similar size as
the electroweak contributions. The recent Brookhaven results on
$a_\mu$ where the first to be sensitive at the level of the
electroweak corrections, and the 2.6 standard deviation discrepancy
with respect to the SM expectations they reported then clearly attracted a lot of attention
from the supersymmetric community\footnote{New theoretical estimates
have changed a sign in the contribution involving photon photon
scattering, reducing this discrepancy to less than two standard
deviations \cite{kn01}.}.   
The magnetic moment has associated with it a flip in the chirality of
 the fermion,
and hence this is why the anomalous magnetic moment of the electron,
although measured more accurately, is not a good place to look for
supersymmetry ($a_e^{SUSY}\sim (m_e/m_\mu)^2a_\mu^{SUSY}$). The
left-right transition in $a_\mu^{SUSY}$ can take place also through
the higgsino vertex (proportional to $\lambda_\mu\sim m_\mu\tan\beta$) or
through a L--R smuon mass insertion in the scalar propagator
(proportional to $m^2_{\tilde\mu LR}=m_\mu(A-\mu\tan\beta)$). Hence
the SUSY contribution is relevant for large tan$\beta$, with typical
predictions being $|a_\mu(SUSY)|\leq 14\times 10^{-10}(100{\rm
GeV}/\tilde m)^2\tan\beta$. 
The reported results (after correcting with the new theoretical
estimates \cite{kn01}), $a_\mu(BNL)-a_\mu(SM)=25(16)\times
10^{-10}$, then suggest that tan$\beta$ should be large and that the
supersymmetric mass scale $\tilde m$ should be below a few hundred
GeV, with two superpartners (one of them being charged) being then at
the reach of future collider experiments. The supersymmetric
explanation of $a_\mu(BNL)$ also preferres positive values of $\mu$,
and this goes in the same direction as $b\to s\gamma$ results, and as
we saw is also good news for future experiments looking for
supersymmetric dark matter. 

\bigskip

In conclusion, the SM has been extremely successful in accounting for
existing experimental data, and has been subject to several  
precision tests. With the energy frontier being pushed above the
electroweak scale by upcoming experiments, and rare processes becoming
tested with high sensitivity, we are entering the stage in which the
search for new physics is becoming the main experimental goal in
particle physics. Certainly some revolution in our understanding of
the underlying laws of nature will come out from this.

\end{itemize}
\section*{BIBLIOGRAPHY}
The present  `light' notes on  the basic aspects of some of 
the major attempts to extend the SM  have the aim to just
 motivate the (experimental and
theoretical) students to undertake the study of this very broad
subject, about which many books and specialized review articles have
been written. Here below are listed  a few of them which can be useful
for this purpose, and in them also a
comprehensive list of references to the original works can be found:
\begin{itemize}
\item P. Langacker, ``Grand Unified Theories and proton decay'', Phys. Rept. {\bf 72}
(1981) 185.

\item G. G. Ross, ``Grand Unified Theories'', Adison-Wesley (1985).

\item J. Wess and J. Bagger, ``Supersymmetry and Supergravity'',
Princeton Series in Physics (1992).

\item H. P. Nilles, ``Supersymmetry, supergravity and particle
physics'', Phys. Rept. {\bf 110} (1984) 1.

\item H. E. Haber and G. Kane, ``The search for supersymmetry: probing
physics beyond the standard model'', Phys. Rept. {\bf 117} (1985) 75.

\item S. Weinberg, ``The Quantum Theory of Fields, part III,
Supersymmetry'', 
Cambridge U. Press (2000).

\item G.F. Giudice and R. Rattazzi, ``Theories with gauge mediated
supersymmetry breaking'', Phys. Rept. {\bf 322} (1999) 419, {\tt
hep-ph/9801271}. 

\item P. Ramond, ``Journeys Beyond the Standard Model'', Perseus Books
(1999). 

\item S. Abel et al., ``Report of the SUGRA working group for run II
of the Tevatron'', {\tt hep-ph/0003154}.

\item N. Polonsky, ``Supersymmetry structure and phenomena'', {\tt
hep-ph/0108236}.

\end{itemize}

\section*{ACKNOWLEDGEMENTS}

I would like to thank the students and the conference organizers for
having made possible this very interesting school. Many thanks also to Gian
Giudice for comments on the manuscript. This work was
partially supported by Fundaci\'on Antorchas.

\end{document}